\documentclass[11pt, a4paper]{article} 
\usepackage{slashed,jheppub,multirow,relsize,soul}
\usepackage[normalem]{ulem}
\usepackage{color}
\usepackage{subcaption}
\usepackage{mathabx}
\usepackage[utf8]{inputenc} 
\usepackage[toc,page]{appendix}
\usepackage{etoolbox}

\usepackage{array}
\newcolumntype{P}[1]{>{\centering\arraybackslash}p{#1}}
\newcolumntype{M}[1]{>{\centering\arraybackslash}m{#1}}

\newcommand{\refeq}[1]{Eq.~(\ref{#1})}

\newcommand{\reffig}[1]{Fig.~\ref{#1}}

\newcommand{\refsec}[1]{Section~\ref{#1}}
\newcommand{\refapp}[1]{Appendix~\ref{#1}}
\newcommand{\reftab}[1]{Table~\ref{#1}}
\newcommand{\refref}[1]{Ref.~\cite{#1}}

\def\lagrangian{lagrangian}
\def\eg{\emph{e.g.}}

\def\muboone{MicroBooNE}
\def\miniboone{MiniBooNE}
\def\icarus{Icarus}
\def\minerva{MINER$\nu$A}
\def\ster{\ensuremath N}

\appto\appendix{\addtocontents{toc}{\protect\setcounter{tocdepth}{1}}}

\title{MeV-scale sterile neutrino decays at the Fermilab Short-Baseline Neutrino program}

\author{Peter Ballett,}
\author{Silvia Pascoli}
\author{and Mark Ross-Lonergan}

\affiliation{Institute for Particle Physics Phenomenology, Department of
Physics, Durham University, South Road, Durham DH1 3LE, United Kingdom}

\emailAdd{peter.ballett@durham.ac.uk}
\emailAdd{silvia.pascoli@durham.ac.uk}
\emailAdd{mark.ross-lonergan@durham.ac.uk}

\preprint{IPPP/16/89}

\abstract{
Nearly-sterile neutrinos with masses in the MeV range and below would be
produced in the beam of the Short-Baseline Neutrino (SBN) program at Fermilab.
In this article, we study the potential for SBN to discover these
particles through their subsequent decays in its detectors.  We discuss the
decays which will be visible at SBN in a minimal and non-minimal extension of
the Standard Model, and perform simulations to compute the
parameter space constraints which could be placed in the absence of a signal.
We demonstrate that the SBN programme can extend existing bounds on well
constrained channels such as $\ster \rightarrow \nu l^+ l^-$ and $\ster
\rightarrow l^\pm \pi^\mp$ while, thanks to the strong particle identification
capabilities of liquid-Argon technology, also place bounds on often neglected
channels such as $\ster \rightarrow \nu\gamma$ and $\ster \rightarrow \nu
\pi^0$. 
Furthermore, we consider the phenomenological impact of improved
event timing information at the three detectors. As well as considering its
role in background reduction, we note that if the light-detection systems in
SBND and ICARUS can achieve nanosecond timing resolution, the effect of finite
sterile neutrino mass could be directly observable, providing a smoking-gun
signature for this class of models.
We stress throughout that the search for heavy nearly-sterile neutrinos is a
complementary new physics analysis to the search for eV-scale oscillations, and
would extend the BSM programme of SBN while requiring no beam or detector
modifications.
}

\begin{document} 

\maketitle

\section{Introduction}

The neutrino sector of the Standard Model (SM) is known to be incomplete. The
observation of oscillatory behaviour between neutrino flavour states
\cite{Fukuda:1998mi} suggests that neutrinos possess a mass matrix with
off-diagonal terms in the flavour basis. There are many models that have been
invoked in the literature to explain this observation as well as the lightness
of neutrino masses, ranging from the ever popular see-saw mechanisms
\cite{Minkowski:1977sc, GellMann:1980vs, Mohapatra:1979ia} to radiative mass
generation \cite{Zee:1980ai,Babu:1988ki} or even more involved constructions
such as neutrino masses originating from extra-dimensions
\cite{ArkaniHamed:1998vp}.  It will ultimately be the role of phenomenology to
find ways to distinguish between potential candidate models, and explore what
can be deduced about the completion of the neutrino sector from the analysis of
contemporary experiments.
A common, although not necessary, feature in Beyond the SM (BSM) models which
account for neutrino masses is the presence of sterile neutrinos, SM-gauge
singlet fermions which couple to the active neutrinos via Yukawa
interactions\footnote{We focus on mass eigenstates which are nearly sterile but
mix with small angles with the active ones. For simplicity, and following
previous literature, we call them ``sterile neutrinos'' throughout the text.}.
After electroweak symmetry breaking, these particles are coupled bilinearly to
the active neutrino fields, and in the mass basis, we find an extended neutrino
sector including new states with mixing-suppressed gauge interactions. \emph{A
priori} their mass and interaction scales can span many orders of magnitude,
leading to a wide range of distinct observable phenomena.
One of the best known examples is the short-baseline oscillation signature
associated with a sterile neutrino mass around the eV-scale (see \eg\
\refref{Gariazzo:2015rra} for a recent review), which has been invoked to
explain anomalies found at some short-baseline oscillation experiments
\cite{Aguilar:2001ty, Aguilar-Arevalo:2013pmq, AguilarArevalo:2008rc}.
Explaining all data in an economical fashion appears challenging in these
models \cite{Kopp:2013vaa,Conrad:2012qt}, but more results would be needed
before a decision can be made as to their role in the neutrino sector. The
Fermilab SBN \cite{Antonello:2015lea} program was primarily designed to perform
such a conclusive test.

The SBN experiment is comprised of three detectors placed in the Booster
Neutrino Beam (BNB) at different (short) baseline distances: SBND (previously
known as LAr1-ND) at 110~m from the target, \muboone\ at 470~m and ICARUS at
600~m.  All three detectors employ Liquid Argon Time Projection Chamber
(LArTPC) technology \cite{Rubbia:1977} with strong event reconstruction
capabilities allowing for a significantly improved understanding of background
processes compared to predecessor technologies. 
With this design, SBN has been shown to be able to extend the current bounds on light
oscillating sterile neutrinos, thoroughly exploring the eV-scale sterile
neutrino mass region, whilst also pursuing many other physics goals
\cite{Antonello:2015lea}.

In this article, we assess SBN's potential to contribute to the search
for sterile neutrinos, in a manner complementary to the oscillatory analysis.
The new fermions in our study are assumed to have masses around the MeV scale.
These particles are light enough to be produced in neutrino beams via meson
decay, but have masses sufficiently large to prevent oscillatory effects with
the active neutrinos through loss of coherence (see \eg\
\refref{Akhmedov:2009rb}), instead propagating long distances along the
beamline. Due to the presence of mixing they are unstable, and their
subsequent decay products can be observed in neutrino detectors.
We stress that the search for MeV-scale sterile neutrinos is entirely
compatible with the primary goals of SBN, and requires modification of
neither the beam nor detector designs. 

The reconstruction \cite{Church:2013hea, Marshall:2015rfa}, energy resolution
\cite{Sorel:2014rka} and excellent calorimetric particle identification
capabilities of LAr \cite{Antonello:2012hu} technology means the SBN program
provides an ideal scenario to study this ``decay-in-flight'' of sterile
neutrinos.  This technology allows for a high degree of background suppression
on well studied decay modes while also allowing the study of channels which
have been poorly bounded by similar experiments due to large backgrounds and
challenging signals. For example, the differentiation between an electron- or
photon- induced EM shower can be achieved by studying their rate of energy loss
in the first $3$ cm of their ionising track \cite{szelc:2007}. Furthermore, as we
discuss in \refsec{sec:timing}, if a sufficiently good timing resolution of
scintillation light is achieved, the timing structure of markedly sub-luminal
sterile neutrinos can be utilised as both a rejection mechanism for beam
related backgrounds as well as a further aid for model discrimination and mass
measurement.

\begin{table}[t!]
\centering
\begin{tabular}{| l || l | l | l | l |}
	\hline
	& PS-191 & SBND & MicroBooNE & ICARUS \\ \hline \hline
	POT	& $0.86 \times 10^{19}$	& $6.6 \times 10^{20}$	&	$13.2 \times 10^{20}$     &  $6.6 \times 10^{20}$ \\ \hline
	Volume	& $216\text{m}^3$	&	$80\text{m}^3$	&	$62\text{m}^3$	     &   $340\text{m}^3$	\\ \hline
	$\text{Baseline}^{-2}$	& $(128 	\text{m} )^{-2}$	&$(110 \text{m} )^{-2}$	&	$(470 \text{m} )^{-2}$			     & $(600 \text{m} )^{-2}$	  \\ \hline
Ratio/PS-191 & - 	& 38.5 	& 3.3	& 5.5\\ \hline
	S/$\sqrt{B}$ Ratio & - 	& 16.3 	& 1.8	& 1.1\\ \hline
\end{tabular}

\caption{\label{tab:exposure} A comparison of the relative exposure at each SBN
detector compared to PS-191. One would expect all  three SBN detectors to see
increased numbers of events than PS-191 did, with SBND seeing the largest
enhancement of a factor of $38.5$. The final row takes into account the scaling
in masses leading to increased backgrounds, although the achievable
reconstruction of LAr should reduce these significantly.}

\end{table}

We restrict our analysis to sterile neutrino masses below the kaon mass. Kaons
and pions are produced in large numbers at BNB, and their subsequent decays
will generate a flux of sterile neutrinos. In this mass range, the strongest
bounds on sterile neutrinos which mix with electron and muon neutrinos come
from PS-191 \cite{Bernardi:1985ny, Bernardi:1987ek}, a beam dump experiment
which ran at CERN in 1984. 
PS-191 was constructed from a helium filled flash chamber decay region,
followed by interleaved iron plates and EM calorimeters. It was located 128~m
downstream of a beryllium target and $2.3^\circ$ (40 mrads) off-axis, obtained
$0.86 \times 10^{19}$ POT over the course of its run-time, and had a total
detector volume of $6\times3\times12 = 216$ m$^3$. We can estimate the
sensitivity of the three SBN detectors and how they will compare to PS-191 by
estimating the experiments' \emph{exposure}, defined here as POT $\times$ Vol
$\times$ $R^{-2}$. We compare the three detectors to PS-191 in
\reftab{tab:exposure}, which indicates that all detectors of the SBN complex
expect a larger exposure, with SBND seeing the greatest enhancement by a factor
of around $40$. 
In addition to the larger exposure, there is also an enhancement of the
expected decay events at SBN due to its lower beam energy. The sterile
neutrinos at SBN are produced by the 8 GeV BNB beam and have a softer spectrum
than those produced by the 19.2 GeV CERN Proton Synchrotron beam used at
PS-191. As we discuss in more detail in \refsec{sec:decays}, the probability
that the sterile neutrino decays inversely scales with momentum, $1/|P_\ster|$,
and we would therefore expect any BNB detectors to see more events than PS-191
for equivalent neutrino exposures.

However, exposure alone does not dictate sensitivity. PS-191 was purposefully
built to search for decays of heavy fermions. To minimise the background
induced by active neutrino scattering, the total mass of the detector (and
therefore number of target nuclei) was chosen to be small (approximately $20$
ton). Conversely, the SBN detectors were designed to search for neutrino
interactions and thus have significantly larger masses ($112$, $66.6$ and $476$
tons respectively). SBN will not only see a greater number of decay events than
PS-191 but also a greater background for a given exposure. Therefore, the
degree of background reduction will be crucial in determining its ultimate
performance. We return to this issue in \refsec{sec:backgroundestimate}.

The paper is structured as follows. In \refsec{sec:decays} we present an
overview of sterile neutrino decay in minimal and non-minimal models relevant
for beam dump experiments. We then present the details of our simulation in
\refsec{sec:simulation} and show illustrative event spectra for some channels
of interest. In \refsec{sec:sensitivities}, we present and discuss the
exclusion contours that SBN could place on the model in the absence of a
signal. We then study how the event timing information could be used to test
the hypothesis of sterile neutrino decay-in-flight and to help constrain the
particle masses if a positive signal were detected. We make some concluding
remarks in \refsec{sec:conclusions}.

\section{\label{sec:decays}Sterile neutrino production and decay}

The most general renormalizable \lagrangian\ extending the SM to include a new
gauge-singlet fermion $\ster$ is given by
\begin{align}   \mathcal{L}_N = \mathcal{L}_\text{SM} +
\overline{\ster}i\slashed{\partial}\ster + \left(\frac{m_\ster}{2}
\overline{\ster^\text{c}}\ster  + y_\alpha\overline{L}_\alpha H
\ster + \text{h.c.}\right),\label{eq:minimallag} \end{align}
where $N^c=C\overline{N}^\text{T}$ with $C$ denoting the charge-conjugation
matrix, $L_\alpha$ is the SM leptonic doublet of flavour $\alpha$, $y_\alpha$
represents the Yukawa couplings and $m_\ster$ a Majorana mass term for $\ster$.
The extension to multiple new fermions involves promoting $y$ and $m_\ster$ to
matrices with indices for the new states, but will offer no real
phenomenological difference in the following analysis.\footnote{The minimal
single $\ster$ extension does not allow for the observed masses of the
neutrinos, as the mass matrix is rank 1. We assume that an appropriate
extension has been introduced to satisfy neutrino oscillation data while
introducing no new dynamics at the lower energy scales of interest.} Much work
has been done understanding the phenomenology of such novel neutral states,
which varies significantly over their large parameter spaces. 
Lagrangians similar to this have been used in the literature for a wide range
of purposes. If the new particle has a mass around $10^{12}$-$10^{15}$ GeV it
could provide a natural way to suppress the size of active neutrino masses
through the Type I or III see-saw mechanisms \cite{Minkowski:1977sc,
GellMann:1980vs, Mohapatra:1979ia}. A lighter neutral fermion, with a mass
around the keV scale, remains a promising dark matter candidate
\cite{Adhikari:2016bei}. A synthesis of these ideas is found in the so-called
$\nu$MSM which simultaneously can explain dark matter, neutrino masses and
successful baryogenesis \cite{Asaka:2005pn}. 
If we consider sterile neutrinos at even lower energy scales, with
masses at the eV scale or below, these particles can alter the neutrino
oscillation probability, leaving observable signatures at oscillation
experiments. Indeed, such particles have been proposed to alleviate
short-baseline oscillation anomalies; although, no minimal solution seems to
provide a compelling universal improvement to the current data
\cite{Kopp:2013vaa,Conrad:2012qt}.

A key feature of models of sterile neutrinos are the weaker-than-weak
interactions which arise from mass mixing. In the minimal \lagrangian\ in
\refeq{eq:minimallag}, the only direct couplings to new sterile flavour
eigenstates are neutrino--Higgs interactions. However, these couplings generate
off-diagonal neutrino bilinears below the electroweak symmetry breaking scale,
leading to mixing-mediated interactions with SM gauge bosons for the mostly
neutral mass eigenstate. This allows them to be produced in and decay via SM
gauge interactions, albeit suppressed by the mixing angle. 

The possibility remains that extra particles exist beyond the minimal
\lagrangian\ and these mediate other interactions, either directly with SM
fields or, as before, via mixing. 
Throughout our work, we assume that the production of $N$, described in
\refsec{sec:prod}, is generated by the interactions in \refeq{eq:minimallag}.
However, we will return to the idea of a non-minimal \lagrangian\ in
\refsec{sec:nonminimal} when discussing the decay modes of $N$.

\subsection{\label{sec:prod}Production at BNB}

For sterile neutrinos which are light enough to be produced from a meson beam,
there is a qualitative divide in the phenomenology somewhere between keV and eV
masses\footnote{The precise mass range depends on details of the process under
consideration.}. If the sterile neutrinos are massive enough for their
mass-splittings with the light neutrinos to be larger than the wavepacket
energy-uncertainty associated with the production mechanism, they no longer
oscillate \cite{Akhmedov:2009rb}.  
Neutral particles produced in the beam will propagate towards the detector and
may be observed by their subsequent decay into SM particles. Experiments
seeking to measure such decays are generally known as beam dump experiments,
where proton collisions with a target produce particles to be observed
down-wind of the source \cite{CooperSarkar:1985nh, Bergsma:1985is,
Vaitaitis:1999wq, Bernardi:1985ny, Bernardi:1987ek, Anelli:2015pba,
Alekhin:2015byh}. It has been pointed out that the difference between a beam
dump and a conventional neutrino beam is more a matter of philosophy than
design, and we can expect many experiments to have some sensitivity to novel
heavy states \cite{Gorbunov:2007ak, Asaka:2012bb, Adams:2013qkq}. 
For the BNB, we can estimate the mass at which the oscillatory behaviour is
suppressed as follows: the decay pipe for BNB is around $50$~m in length, which
is considerably shorter than the decay lengths of the mesons in the beam, and
we assume that this length defines the wavepacket width at production.  The
relevant parameter is the decoherence parameter \cite{Akhmedov:2009rb,
Hernandez:2011rs}
$\xi = 2\pi \frac{\lambda_\text{d}}{\lambda_\nu},$
where $\lambda_\text{d} = 50$~m and $\lambda_\nu$ is the standard neutrino
oscillation length $\lambda_\nu = \Delta m^2/4E_\nu$. For $\xi\gg1$ the wave
packet is insufficiently broad to accommodate a coherent superposition of the
heavy and light neutrino states. We estimate that this occurs for the BNB at 
$ \Delta m^2 \gtrsim 100~\text{eV}^2.$

In a conventional neutrino beam, most neutrinos are derived from meson decay,
and we assume in this work that the sterile neutrinos are produced from the
decays of pions and kaons, restricting our sterile neutrino mass to $m_\ster
\le m_K$.
Larger sterile neutrino masses could be probed by working at higher energies in
the initial proton beam, where the neutral fermions could come from decays of
charmed mesons such as $D^\pm$. This strategy will be used by the upcoming
SHiPS experiment \cite{Alekhin:2015byh, Anelli:2015pba} but will not be
considered further in the present work as $D$ mesons are produced in extremely
small numbers due to the relatively low energy of the BNB beam
\cite{AguilarArevalo:2008yp}. As such we restrict ourselves to the naturally
defined mass range of interest for SBN, eV $\ll m_\ster \lesssim$
494~MeV.  We focus on $m_\ster \gtrsim$ MeV scale states where the prospects
for detection are greatest due to enhanced decay rates.

Although novel dynamics may lead to enhanced production rates of sterile
neutrinos by alternative unconventional means, we neglect this possibility and
assume that the sterile component of the BNB flux arises solely from meson (or
secondary $\mu^\pm$) decays. This process requires only mass-mixing from the
$N$-$\nu$ Yukawa term in \refeq{eq:minimallag}. It follows that the amplitudes
for these decays are related to those of the standard leptonic decays of mesons
via an insertion of the mixing matrix element $U_{\alpha 4}$, and to leading
order in the mass of the sterile neutrino over the meson mass, the
$\ster$-fluxes will be a rescaling of the fluxes for the active neutrinos. 
However in order to account for flavour-specific effects, it is necessary to go
beyond this approximation and consider the kinematic differences of heavy
sterile neutrino production.
The flux of sterile neutrinos produced from the decay of a given meson $M$ is
approximated by
\begin{equation} \phi_{\ster}(E_{\ster}) \approx \phi_{\nu_\alpha} (E_{\nu_\alpha})\vert
U_{\alpha 4}\vert^2 \frac{\rho\left( \delta_M^a , \delta_M^i
\right)}{\delta_M^a \left(1- \delta_M^a\right)^2}, \label{eq:flux_approx} \end{equation}
where $\rho(a,b)=\mathcal{F}_M(a,b) \lambda^{\frac{1}{2}}(1,a,b)$ is a
kinematic factor consisting of a term proportional to the two body phase space
factor, $\lambda(x,y,z)=x^2+y^2+z^2-2(x y+yz+x z)$ and a term proportional to
the matrix element, $\mathcal{F}_M(a,b)= a+b -\left(a-b\right)^2$, with
$\delta_M^{a(i)}=m_{l_a(\nu_i)}^2/M^2$ \cite{PhysRevD.24.1232}. 

The kinematic factor leads to two effects. First, it provides a threshold
effect of suppressing the production when the phase space decreases near a
kinematic boundary.  Secondly, it allows for the helicity un-suppression of
channels which in a conventional beam are highly suppressed. For example, the
decay $\pi^\pm \to e^\pm \nu_e$ which is significantly suppressed compared to
the muonic channel, sees no such suppression when the neutrino is replaced with
$N$.
This kinematic effect for the pion and kaon can be substantial, for $\pi
\rightarrow e \nu$ this factor can be as large as $10^5$, which more than
compensates for the significantly smaller intrinsic flux of $\nu_e$
intrinsic to the BNB, which is around $0.52$\% of the total flux
\cite{AguilarArevalo:2008yp}. The approximation in \refeq{eq:flux_approx}
starts to fail as the mass of the sterile neutrino increases, and we begin to
see components of the active flux having energies less than the sterile mass
which are truncated by the kinematic factor. In order to keep the normalisation
of total neutrino events constant, before $U_{\alpha 4}$ and kinematic scaling,
any events which are below the sterile neutrino mass threshold are removed and
the remaining flux is scaled accordingly.

\begin{figure}[t]
\centering
\begin{subfigure}{.5\textwidth}
  \centering
  \includegraphics[width=\linewidth]{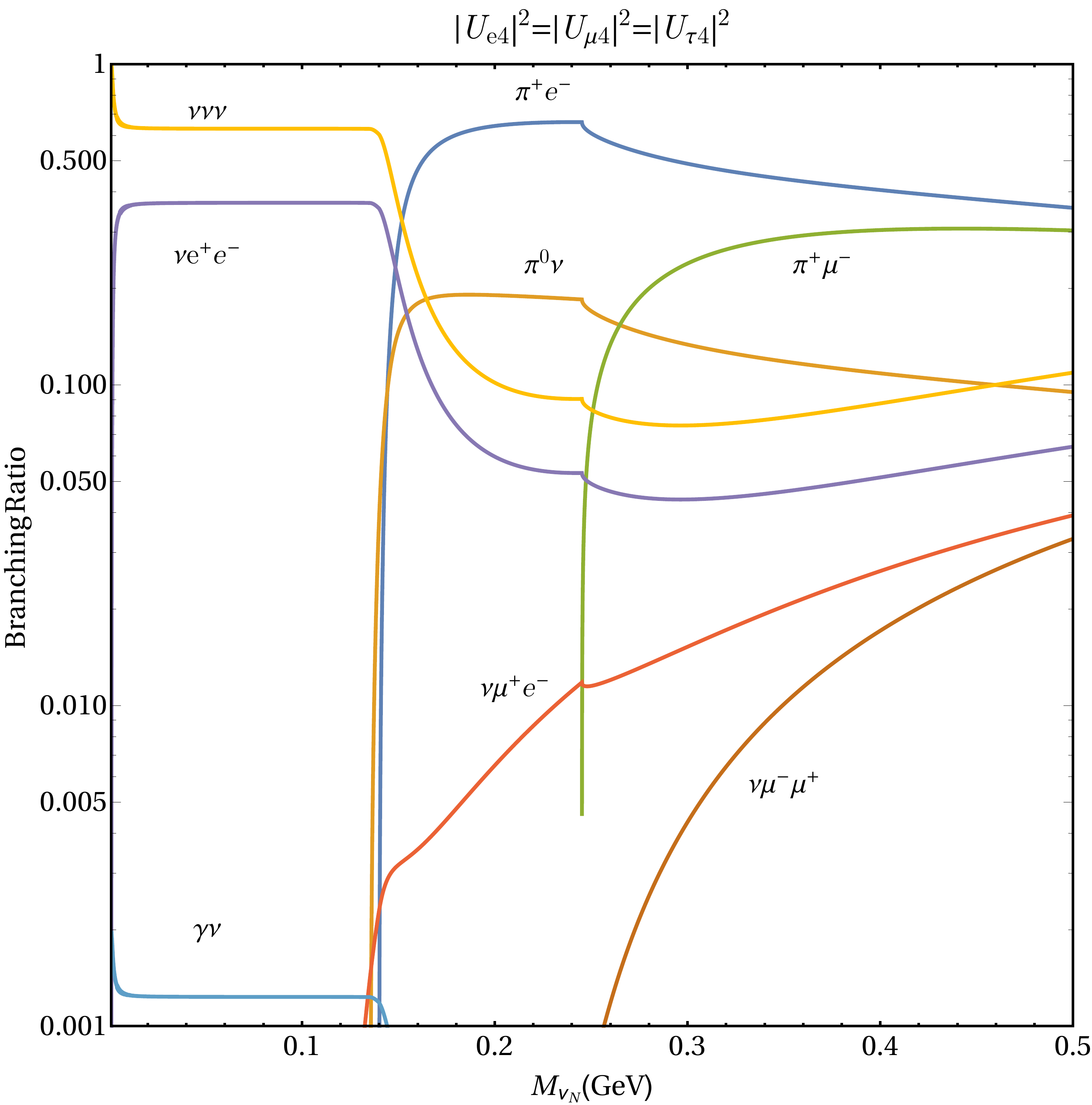}
\end{subfigure}%
\begin{subfigure}{.5\textwidth}
 \centering
\includegraphics[width=\linewidth]{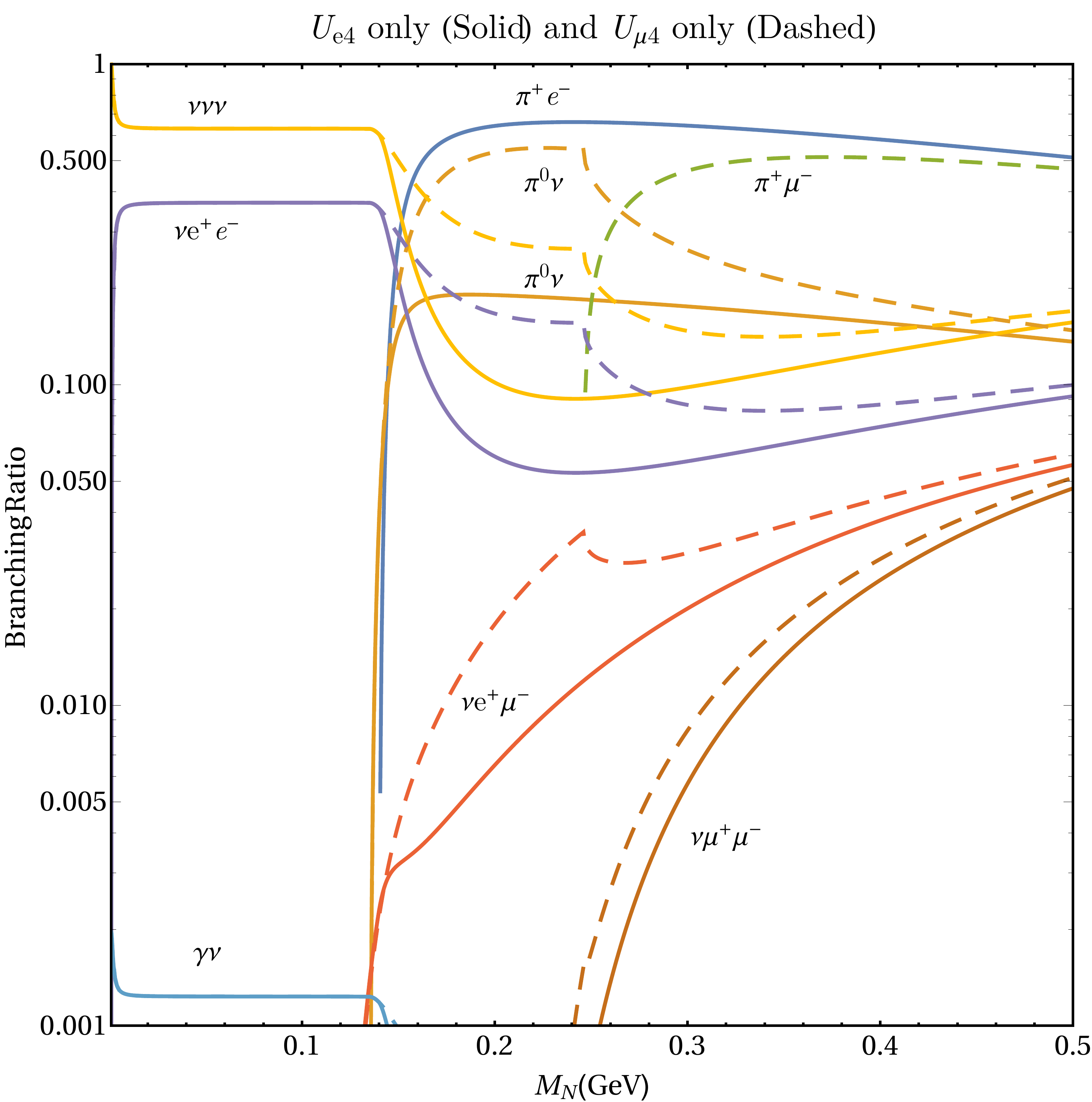}
\end{subfigure}

\caption{\label{fig:branchingratios}The branching ratios for sterile neutrino
decays in the minimal 3 sterile neutrino SM extension, with masses between 1
MeV and 0.5 GeV. We assume both a flavour independent mixing pattern (left
panel) and a hierarchical scenario (right panel) in which either $U_{e4}$
(solid lines) or $U_{\mu 4}$ (dashed lines) is the dominant mixing-matrix
element.}

\end{figure}

\subsection{Decay at SBN}

The fermions $N$ will generally be unstable, albeit possibly long-lived,
allowing for decays-in-flight into SM particles. In this work, we try to keep
an open mind about the interactions of the sterile neutrino and consider all
kinematically possible tree-level decays to visible SM particles for sterile
fermions produced by pion and kaon decays, $10~\text{MeV}\lesssim m_{\ster}
\lesssim m_K$. The precise decay rates and branching ratios for these channels
are model dependent. In this section, we discuss the decay rates
for a minimal extension of the SM, as well as the implications of a non-minimal
model.

\subsubsection{\label{sec:minimal}Minimal model}

We define the minimal sterile neutrino model by the Lagrangian in
\refeq{eq:minimallag}. This encompasses the best known model of sterile
neutrino phenomenology --- the UV-complete Type I see-saw (and its low-scale
variants) --- but also provides an effective description of more complicated 
extensions of the SM in which the additional field content does not directly 
affect the neutrino sector at low energies.
Decays of sterile neutrinos in such a model proceed via SM interactions
suppressed by the mixing angle and have been studied in \refref{Atre:2009rg,
PhysRevD.24.1232, PhysRevD.24.1275}. We have plotted the branching ratios
for sterile neutrinos in our mass range in \reffig{fig:branchingratios}, and 
we will now briefly summarise the decay rates most important for the present
study.

The decays of the minimal model depend only on the mass of the $\ster$ and the
size of neutrino mixing to various flavours, parameterized by the elements of
an extended PMNS matrix, \emph{e.g.} for one additional particle 
$U_{\alpha 4}$ for $\alpha \in \{e,\mu,\tau\}$. 
The branching ratios for these decays are shown in \reffig{fig:branchingratios}
as a function of mass for two sets of assumptions about the PMNS matrix. On the
left, we show the branching ratios if all mixing elements are of a similar
size, whereas on the right we assume that only $U_{\mu4}$ or $U_{e4}$ are
non-zero. This leads to certain semi-leptonic decays being forbidden,
significantly changing the phenomenology of the model for some masses.

For sterile neutrino masses less than the pion mass, the dominant decay is into
three light neutrinos. This channel is for all practical purposes unobservable
and we will not consider it further. The dominant decay into \emph{visible}
particles will be into an electron-positron pair with a branching fraction of
around $38\%$. 
This is true regardless of the flavour structure of the mixing matrix;
although, this decay channel is not flavour-blind. If the sterile neutrino
mixes primarily through $U_{e4}$, the decay proceeds via both neutral and
charged currents, but for $U_{e4}=0$, this channel occurs via neutral current
only. The decay rate for this channel is given by 
\begin{align*} \Gamma\left(\ster\to \nu_\alpha e^+e^-\right) =
\frac{G_\text{F}^2m_\ster^5}{96\pi^3}\left|U_{\alpha 4}\right|^2&\left[\left(
g_Lg_R + \delta_{\alpha e}g_R\right)I_1\left(0,\frac{m_e}{m_\ster}, \frac{
m_e}{m_\ster}\right)\right.\\ &\left.\qquad + \left(g_L^2 + g_R^2 +
\delta_{\alpha
e}(1+2g_L)\right)I_2\left(0,\frac{m_e}{m_\ster},\frac{m_e}{m_\ster}\right)\right],
\end{align*}
where $I_1(x,y,z)$ and $I_2(x,y,z)$ are integrals over phase space such that
$I_1(0,0,0) = 1$ and $I_2(0,0,0) = 0$. Further details of the decay rates used
in this work are given in \refapp{app:decayrates}.
Although the electron-positron channel dominates the visible decays at $m_\ster
\leq m_\pi^0$, we also consider the radiative decay $\ster\to\nu_i\gamma$ which
would generate an observationally challenging single photon signal
\cite{PhysRevD.25.766}. In the minimal model the decay occurs via a
charged-lepton/$W$ loop and has a rate given by
\[ \Gamma(\ster\to\nu_i\gamma) = \frac{G_\text{F}^2m_\ster^5 |U_{\alpha
4}|^2}{192 \pi^3} \left( \frac{27 \alpha}{32 \pi} \right). \]
This decay channel is significantly suppressed by the light mass of the sterile
neutrino, the mixing-matrix elements and the loop factor. It can
be estimated at around $\Gamma(\ster\to\nu_i\gamma)/(\text{GeV}) \approx
10^{-20} (m_\ster/\text{GeV})^5$. We see in \reffig{fig:branchingratios} that
this leads to a branching ratio of around $10^{-3}$.

Additional leptonic decays open up for sterile neutrino masses which satisfy
$m_\ster \geq m_\mu+m_e$. Although with a smaller branching ratio, decays
involving muons are clean signatures at LAr detectors. In the case of
$N\rightarrow \nu_\alpha \mu^+ \mu^-$ the decay occurs by both neutral and
charged currents and follows from the $N\rightarrow \nu_\alpha e^+ e^-$ decay
given above with the replacement $m_e \rightarrow m_\mu$.  The mixed-flavour
decays, \emph{e.g.} $N\rightarrow \nu_\alpha \mu^\pm e^\pm$, occur by charged
current only and are given by 
\[ \Gamma(\ster\to\nu_{\alpha} \beta^- \alpha^+) = \frac{G_\text{F}^2m_\ster^5
|U_{\beta 4}|^2}{192 \pi^3} I_1\left(\frac{m_\beta}{m_N},\frac{ m_\alpha}{m_N},
\frac{ m_\alpha}{m_N} \right) , \] 
with $\{\alpha,\beta\} = \{e, \mu\}$. The next thresholds lie just above the
pion mass, where two further decays become possible: $N\to\nu \pi^0$ and $N\to
e^\pm\pi^\mp$. These processes quickly become the dominant decays at this mass
range. 
The decay rate for the first process is given by
\[ \Gamma\left(\ster \to \nu_i \pi^0\right) = \sum_{\alpha}
\frac{G_\text{F}^2f_\pi^2m_\ster^3\left|U_{\alpha 4}\right|^2}{64\pi}
\left[1-\left(\frac{m_\pi}{m_\ster} \right)^2\right].  \]
The decay into a charged pion and a lepton is an important channel, and one of
the most constrained in direct decay experiments due to its clean
two-body signal. Its decay rate has a similar algebraic form to the rate of
$N\to \nu \pi^0$ with the addition of a CKM matrix element arising from the
$W$-vertex,
\begin{align} \Gamma\left(\ster\to l^\pm\pi^\mp\right) =
\left|U_{l4}\right|^2\frac{G_\text{F}^2f_\pi^2 |V_{ud}|^2
m_\ster^3}{16\pi}I\left(\frac{m_l^2}{m_\ster^2} ,
\frac{m_\pi^2}{m_\ster^2}\right) , \label{eq:chargedlep_decayrate}\end{align}
where $I(x,y)$ is a kinematic function which away from the production threshold
provides a small suppression factor of around $0.5$. Further
details are given in \refapp{app:decayrates}.
If it is allowed by the flavour structure, the $N\to e^\pm\pi^\mp$
channel dominates the branching ratios for sterile neutrino masses which
satisfy $m_{\pi^\pm} \lesssim m_\ster$. However, as it is mediated by a $W$
boson, in the absence of $U_{e4}$ mixing, this decay would be forbidden and the
decay into a neutral $\pi^0$ and a light neutrino would be dominant. Once the
mass of the sterile fermion is above $m_\ster \gtrsim 235$ MeV, the
$\mu^\pm \pi^\mp$ charged-lepton pion decay opens up. This is another strongly
constrained channel, and its decay rate is already given in
\refeq{eq:chargedlep_decayrate} with $m_l = m_\mu$. Although this decay rate
can also be arbitrarily suppressed by reducing the size of $U_{\mu 4}$, due to
the constraint that all sterile neutrinos must be produced through $U_{\mu 4}$
or $U_{e4}$ mixing, in no case will both of the $l^\pm\pi^\mp$ channels be
absent. As can be seen in the right panel of \reffig{fig:branchingratios}, we
can expect one of them to dominate for higher masses.

\subsubsection{\label{sec:nonminimal}Non-minimal models}

In the previous section we discussed the decay rates for the minimal model of
\refeq{eq:minimallag}. Although such low-scale see-saw models provide a viable
and phenomenologically interesting region of parameter space for both masses
and mixing, they lack a theoretically appealing mechanism to explain the
sub-electroweak sterile neutrino mass scale and the sizes of active neutrino
masses.
Alternative models exist which feature light neutral particles,
but these rely on additional fields or symmetries to help explain these
scales.
Indeed it has been stressed before \cite{delAguila:2008ir} that the discovery
of a light sterile neutrino would necessitate not just the addition of new
neutral fermions to the SM but would be a sign of the existence of some
non-trivial new physics with which to stabilise the mass scale.
 
If heavy novel fields are present in the full model, we can view
\refeq{eq:minimallag} as the renormalizable terms of an effective \lagrangian.
The effective field theory of a SM extension involving new sterile fermions has
been considered at dimension 5 \cite{delAguila:2008ir,Aparici:2009fh},
dimension 6 \cite{delAguila:2008ir} and dimension 7
\cite{Bhattacharya:2015vja}.
We extend the field content of the SM by a sterile fermion $N$\footnote{As
before, we focus on the single $N$ case for illustrative purposes. The
extension to a set of fields $N_i$ is straightforward.}.  The \lagrangian\ can
then be decomposed as a formal power series of terms of increasing dimension
$d$, suppressed by a new physics energy scale $\Lambda$,
\[  \mathcal{L} = \mathcal{L}_N + \sum_{d=5}^\infty
\frac{1}{\Lambda^{d-4}}\mathcal{L}_d, \]
where $\mathcal{L}_N$ is given by \refeq{eq:minimallag} as the sum of the SM
and renormalizable terms including $N$. In \refref{Aparici:2009fh} the
phenomenology of the effective operators at dimension 5 are considered in
detail. Along with the Weinberg operator, which could be the source of a light
neutrino Majorana mass term \cite{Weinberg:1979sa}, the authors find two
effective operators: an operator coupling the sterile neutrino to the Higgs
doublet and a tensorial coupling between the sterile neutrino and the
hypercharge field strength 
\[ \mathcal{L}_5 \supset \frac{c_1}{\Lambda}\overline{\ster^c}\ster(H^\dagger
H) + \frac{c_2}{\Lambda}\overline{\ster^c}\sigma^{\mu\nu}\ster B_{\mu\nu}. \] 
At energies below the electroweak scale, and after diagonalisation into mass
eigenstates for the neutrinos, these operators generate novel couplings, for
example a vertex allowing $\ster\to h \nu$ ($\ster_1\to h \ster_2$), $\ster\to
\nu Z$ ($\ster_1\to Z \ster_2$) and $\ster \to \nu \gamma$ ($\ster_1 \to
\ster_2 \gamma$) at a rate governed by the scale of new physics suppressing
these operators.
Of particular interest is the electroweak tensorial operator, which induces a
rich range of phenomena \cite{Aparici:2009fh}. In the mass range of interest in
the present work, bounds on such an operator are fairly weak: strong
constraints from anomalous cooling mechanisms in astrophysical settings apply
only for lower sterile neutrino masses, whilst collider bounds only become
competitive for higher masses. This could also be related to the enhanced
$\ster \to \nu\gamma$ decay rate introduced in
\refref{Gninenko:2009ks,Gninenko:2010pr} to explain the short-baseline
anomalous excesses.
See also \refref{Duarte:2016miz} for a discussion of decay rates in the
effective sterile neutrino extension up to dimension 6.

If light degrees of freedom are present in addition to (or instead of) heavy
ones, the predictions could be very different from those derived from the
minimal model or the low-energy effective theory.
For example, models with sterile neutrinos that also feature novel interactions
can have significantly different decay rates and branching fractions,
strengthening some bounds and invalidating others
\cite{Batell:2016zod}.
As an example, a model with a leptophilic $Z^\prime$ \cite{Foot:1994vd} could
enhance the magnitudes of some leptonic decay rates, such as $\ster\to \nu e^+\mu^-$, 
while leaving unchanged semileptonic processes like $\ster\to e^\pm \pi^\mp$.
Often, the bounds on masses and mixing angles in these models need to be reconsidered.

For the reasons discussed so far, it is desirable to place bounds on all
possible decays of a neutral fermion allowing for non-standard decay rates to
visible particles. The main consequence of this is that there is \emph{a
priori} no known relationship between the  magnitude of the different decay rates --- a
single channel may be enhanced beyond its value from \refsec{sec:minimal} ---
and bounds inferred from the non-observation of a given channel may not hold in
a non-minimal model when applied to another channel. We therefore do not
restrict our study to those decays which lead to the most stringent bounds on
the parameters of the minimal model, instead studying all kinematically viable
decays independently.

\subsection{\label{sec:bounds}Existing bounds on $U_{\alpha 4}$}

The minimal \lagrangian\ in \refeq{eq:minimallag} has been the basis of many prior
experimental searches for heavy sterile fermions, leading to a variety of
bounds on the magnitude of the active-sterile mixing relevant for sterile
neutrino masses around the MeV-scale. In this section we discuss the relevance
of three key bounds on our model: peak searches, beam dumps and non-terrestrial
considerations.   

An established way to find strong model independent bounds on heavy sterile
neutrinos is through the study of two-body decays of mesons, particularly pions
and kaons \cite{PhysRevD.46.R885, PhysRevLett.68.3000}. Due to the two-body
kinematics, the magnitude of the neutrino mass manifests itself as a
monochromatic line in the charged lepton energy spectrum at $E_l = \left(
m_{\pi (K)}^2+m_l^2-m_N^2\right)/m_{\pi(K)}^2$. These peak searches provide
strong bounds on the sterile-active mixing, while remaining agnostic as to the
ultimate fate of the sterile neutrino, which may be extremely long
lived\footnote{If, on the other hand, the sterile neutrino is
extremely short-lived, these bounds may be weakened. If the particle decays on
the scale of the experiment, it may produce a multi-lepton final state and
escape observation by the single-lepton analysis cuts.}.  Meson decay peak
searches have taken place for $\pi\rightarrow \nu e (\mu)$ and $K \rightarrow
\nu e (\mu)$ and strongly bound active-sterile mixing angles at low masses. The
strength of these bounds is not a function of sterile neutrino decay-rate, and
as such, peak searches tend to perform worse at higher masses in comparison to
bounds from experiments which derive their signal from large sterile neutrino
decay rates.

The tightest bounds on MeV scale sterile neutrinos come from beam dump
experiments. Beam dump experiments study the particles emitted during proton
collisions with a target.  Although BSM particles may be produced
directly \cite{Coloma:2015pih,Dharmapalan:2012xp}, sterile neutrinos would
predominantly arise as secondary decay products of mesons produced in the
initial collision. The set-up required for such an experiment is quite minimal
--- a proton beam, a target and a down-wind detector --- and for this reason
searches of this type have taken place at many accelerator complexes, taking
advantage of preexisting proton beams in their design. Seeking to produce and
observe the subsequent decay of the sterile neutrinos, the sensitivity of beam
dump experiments is driven by both flux intensity and the decay rate of the
heavy sterile neutrino, which typically scales as $(\Gamma \propto m_\ster^3)$
$\Gamma \propto m_\ster^5$ for (semi-) leptonic decays. As such they typically
set tighter bounds as the sterile neutrino mass increases. As discussed in the
introduction, PS-191, which ran in parallel with the BEBC bubble chamber,
provides the strongest limits on active-sterile mixing for masses below the
kaon mass. Above this mass, a higher energy proton beam is needed to further
the same strategy. This was achieved by moving from the CERN PS to
the SPS proton beam in both the CHARM \cite{Bergsma:1985is} and
NA3 \cite{Badier:1986xz} experiments. Beam dumps are incredibly sensitive to
active-sterile mixing and limits $|U_{e 4}|^2 \leq 10^{-8}$--$10^{-9}$ were set
for $m_\ster \geq 200$ MeV.

Results from beam dump experiments are most often presented, as we
did above, as upper limits on active-sterile mixing in the context of the
minimal model. However, beam dump experiments actually set two bounds: there is
also a lower bound on the mixing-matrix elements, where the decay rate is so
large that the sterile neutrino beam attenuates en route to the detector. 
In the minimal model, this lower bound is often at very large values of
$|U_{\alpha 4}|^2$, presenting consistency issues with unitarity data, and is
justifiably ignored. If one considers enhanced decay rates in a non-minimal
model, however, care must be taken with existing bounds as an enhanced decay
rate would modify both bounds. This can reduce the applicability of certain
bounds to non-minimal models. It is instructive to discuss how to scale
existing bounds on the minimal model, or indeed the bounds we will present in
\refsec{sec:sensitivities}, to an extended model which has an
enhancement in the decay rate for one or more channels.
By comparing the flux-folded probabilities to decay inside a
detector for a beam dump experiment of baseline $L$ and detector length
$\lambda$, we can map the published lower bound, $|\widetilde{U}_{\alpha
4}|^2$, to both the new upper and lower bound on the mixing matrix element in a
non-minimal model, $|U_{\alpha 4}|$. For a generic non-minimal model in which
the total decay rate is scaled by a factor $A$ with respect to the minimal
model, and the decay rate into the channel of interest is scaled by a factor
$B$, the constraint takes the form of Lambert's equation (at leading order in
$\lambda/L$), and the bounds on the non-minimal mixing-matrix element are given
by the two real branches of the Lambert-$W$ function,
\[ \frac{|\widetilde{U}_{\alpha 4}|^2}{B \kappa}
\mathcal{W}_{-1}\left(\exp^\kappa  \frac{B}{\sqrt{A} } \kappa  \right) \quad
\leq \quad |U_{\alpha 4}|^2 \quad \leq\quad \frac{|\widetilde{U}_{\alpha
4}|^2}{B \kappa} \mathcal{W}_{0}\left(\exp^\kappa  \frac{B}{\sqrt{A} } \kappa
\right), \label{eq:genericscale} \]
where $\kappa \equiv -\Gamma_\text{T} L / (2 \gamma \beta)$ with
$\Gamma_\text{T}$ the total decay rate calculated with $|\widetilde{U}_{\alpha
4}|^2$.  The upper bound is primarily dependent on the decay rate into the
channel of interest, governed by the parameter $B$, whilst the lower bound is
predominantly sensitive to the total decay rate and the parameter $A$.
Physically, the upper bound is seen to depend on how many decays are produced
and is sensitive to the (possibly enhanced) decay rate into that channel, while
the lower bound arises when the beam attenuates due to decay before the
detector, the rate of which is governed by the total decay rate.

Although the exact behaviour of the bounds for a non-minimal BSM extension are
model-dependent due to correlations between $A$ and $B$, in many situations the
upper bound can be significantly simplified. 
We consider two distinct scenarios depending on whether the enhancement affects
the decay rate of the channel being observed, or another decay channel. We
write the total decay rate as $\Gamma_\text{T} = \Gamma_\text{o} +
\Gamma_\text{c}$, where $\Gamma_\text{c}$ denotes the channel whose decay
products are being measured and $\Gamma_\text{o}$ the sum of all other decay
rates. In our first scenario, the only enhancement is to the channel of
interest, and the total decay rate can be written as $\Gamma_\text{T} =
\Gamma_\text{o}+B\Gamma_\text{c}$. 
In this case, the upper bound from \refeq{eq:genericscale} can
be simplified by expanding in the published bound\footnote{These are typically
of the order $10^{-4}$--$10^{-8}$ and such an expansion is a very good
approximation.}, $|\widetilde{U}_{\alpha 4}|^2$. In this approximation, the new
bound is seen to be a simple scaling of the old bound
\[ |U_{\alpha 4}|^2 \leq \frac{|\widetilde{U}_{\alpha 4}|^2}{ \sqrt{B}}. \] 
This follows our naive expectations: a larger decay rate produces more events
and so bounds are proportionally stronger. The lower bound on $|U_{\alpha
4}|^2$ has no corresponding simple form, but numerically can be seen to follow
a similar scaling relationship: as the enhancement goes up, the bound moves to
lower values of the mixing-matrix element. In this case, apart from a
replacement of the minimal $|\widetilde{U}_{\alpha 4}|^2$ by an effective
mixing-matrix element $|U_{\alpha 4}|^2/\sqrt{B}$, the bounds are to a good
approximation unchanged.
The situation is qualitatively different in our second scenario, however. In
this case, we consider an enhancement to $\Gamma_\text{o}$, so that
$\Gamma_\text{T} = A \Gamma_\text{o}+\Gamma_\text{c}$. We find that the upper
bound is unchanged to leading order, $|U_{\alpha 4}|^2 \leq
|\widetilde{U}_{\alpha 4}|^2$. However, the lower bound moves to smaller values
as the enhancement increases.
For large enhancements, this can significantly reduce the region of parameter
space in which an experiment can bound the model. We will return to these
simplified models of decay rate enhancement in \refsec{sec:sensitivities}.

We note in passing that the limit of large $\lambda/L$ can also be relevant for
\refeq{eq:genericscale}. This corresponds to experiments where production and
detection occur inside the detector, which can be seen as zero baseline
beam-dumps. We find that these experiments produce only an upper bound on the
mixing angle, as the number of incoming sterile neutrinos can no longer be
attenuated through decays occurring before the detector.

Although peak searches and beam dumps set some of the most stringent upper
limits on mixing, non-terrestrial measurements may also place bounds on such
long lived sterile neutrinos due to their effect on the evolution of the early
universe. Heavy sterile neutrinos can have a strong impact on the success of Big-Bang
Nucleosynthesis (BBN) by both speeding up the expansion of the universe with
their additional energy, and thus effecting an earlier freeze out of the
neutron-proton ratio, as well as potentially modifying the spectrum of active
neutrinos via their subsequent decays. If, however, the sterile has
sufficiently short lifetime then their effect on BBN is mitigated as the bulk
of thermally produced sterile neutrinos have decayed long before $T_\text{BBN} \approx
10$ MeV \cite{Fields:2006ga}. The strength of these bounds have been estimated
conservatively for a single sterile neutrino, $m_N < m_{\pi^0}$, as
\cite{Dolgov:2000jw,Dolgov:2000pj}
\begin{align*} \tau_\text{N} &< 1.287 \left(
\frac{m_N}{\text{MeV}}\right)^{-1.828}+0.04179 \text{  s    $\qquad$  for
$U_{\mu 4}$ or $U_{\tau 4}$ mixing},\\ \tau_\text{N} &< 1699 \left(
\frac{m_N}{\text{MeV}}\right)^{-2.652}+0.0544 \text{  s    $\qquad$  for $U_{e
4}$ mixing}, \end{align*}
at the 90\% CL. Although the scenario for $m_N > m_{\pi^0}$ has not been
studied in as much detail, an often quoted bound is that $\tau_N > 0.1$~s is
excluded under current BBN measurements \cite{Dolgov:2000jw}. In the minimal
model, this upper bound on the sterile neutrino lifetime is directly mapped to
a minimum bound on the active-sterile mixing elements $U_{\alpha 4}$. However,
even a modest increase in the total sterile neutrino decay rate, for example by
additional interactions in the sterile neutrino sector leading to decays that
are not mixing suppressed, pushes the total sterile neutrino lifetime below
$0.1$ s and avoids these bounds, while still leaving channel specific
signatures observable at SBN as the upper bounds are independent on total decay
rate. Similarly in a non-standard model of the early universe, these bounds may
not apply.  Therefore, although setting important complementary bounds on
models of sterile neutrino decay, model dependent factors make it possible for
discrepancy between peak search, beam dump and cosmological constraints. As
such a wide program of experimental work is desirable, with as varied a
methodology as possible, to best identify new physics.

\section{\label{sec:simulation}Simulation of SBN}

SBN consists of three LArTPC detectors (SBND, \muboone\ and ICARUS) located in
the Booster neutrino beam. The Booster neutrino beam is a well understood beam,
having been recently studied by the \miniboone\ experiment. For the purposes of
this analysis each detector is assumed to be identical apart from their
geometric dimensions. We simulate the event numbers and distributions at each
detector site using a custom Monte Carlo program which allows efficiencies to
be taken into account arising from experimental details such as energy and
timing resolution in a fully correlated way between observables, and provides
us with event level variables for use in a cut-based analysis. We compute the
total number of accepted events in channel ``$\text{c}$'' via the following
summation,
\[ \ster_\text{c} = \sum_{i} \left .
\frac{\mathrm{d}\phi}{\mathrm{d}E}\right|_{E_i} P_\text{D}\left(E_i\right)
W_\text{c}\left(E_i\right),  \]
where $P_\text{D}(E)$ is the probability for a sterile neutrino of energy $E_i$
to travel the baseline distance and then decay inside the detector labelled
$\text{D}$. The simplest approximation is to ignore all geometric effects, so
that every particle travels exactly along the direction of the beam line, which
gives the following probability 
\[ P_\text{D}\left(E\right) = e^{-\frac{\Gamma_\text{T}L}{\gamma\beta}}\left(
1-
e^{-\frac{\Gamma_\text{T}\lambda}{\gamma\beta}}\right)\frac{\Gamma_\text{c}}{\Gamma_\text{T}},
\label{eq:prob} \]
where $\Gamma_\text{T}$ ($\Gamma_\text{c}$) denotes the rest-frame total decay
width (decay width into channel $\text{c}$), and $L$ ($\lambda$) the distance
to (width of) the detector. The combination $\gamma\beta$ is the usual special
relativistic function of velocities of the parent particle and provides the
sole dependence on energy and sterile neutrino mass $m_\ster$ of the expression
\[   \frac{1}{\gamma\beta} \equiv \frac{m_\ster}{\sqrt{E^2-m_\ster^2}}. \]

\begin{figure}[t]
\center
\begin{subfigure}[t]{0.5\textwidth}
\includegraphics[width=\textwidth]{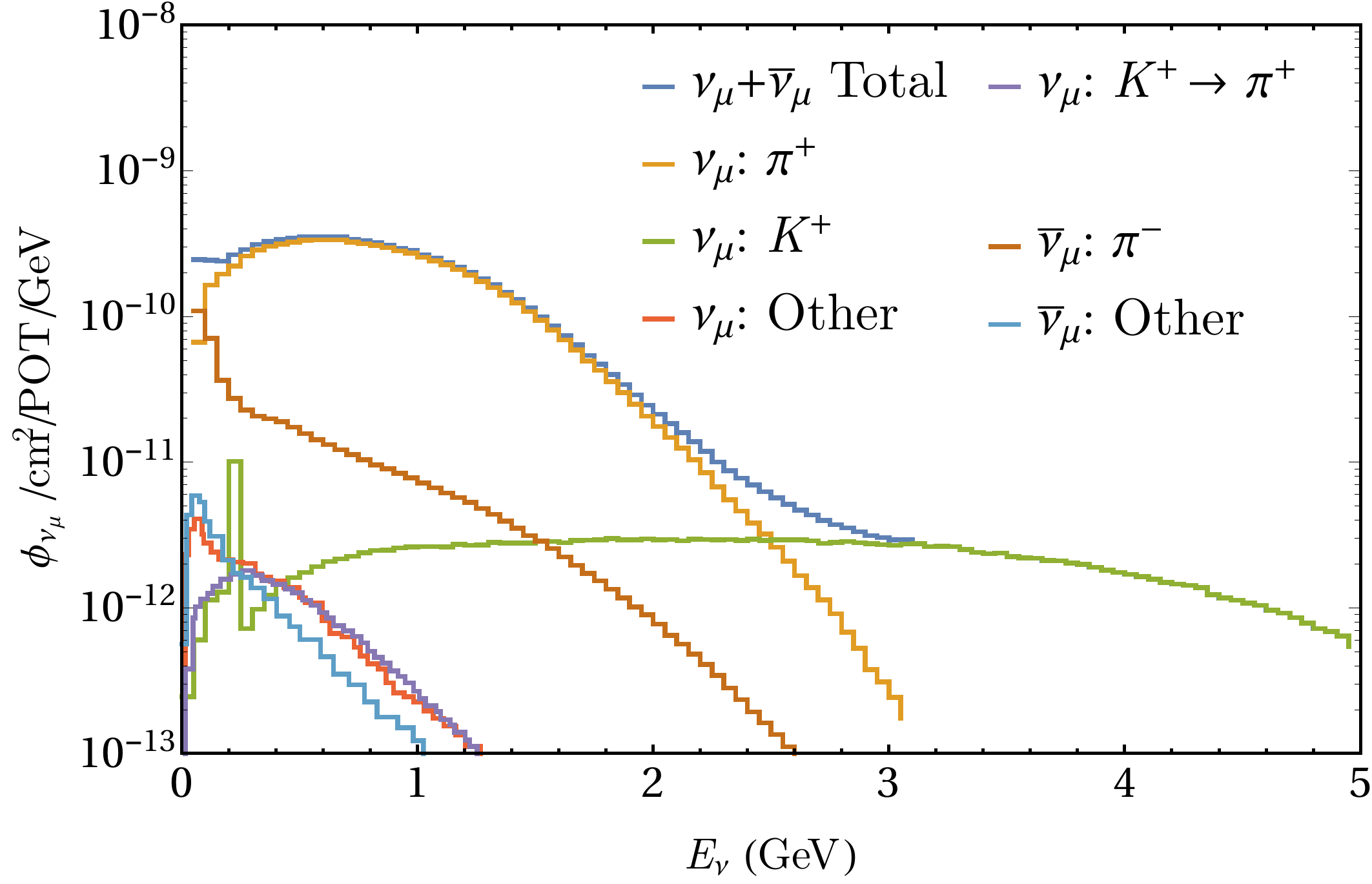} 
\end{subfigure}%
~
\begin{subfigure}[t]{0.5\textwidth}
\includegraphics[width=\textwidth]{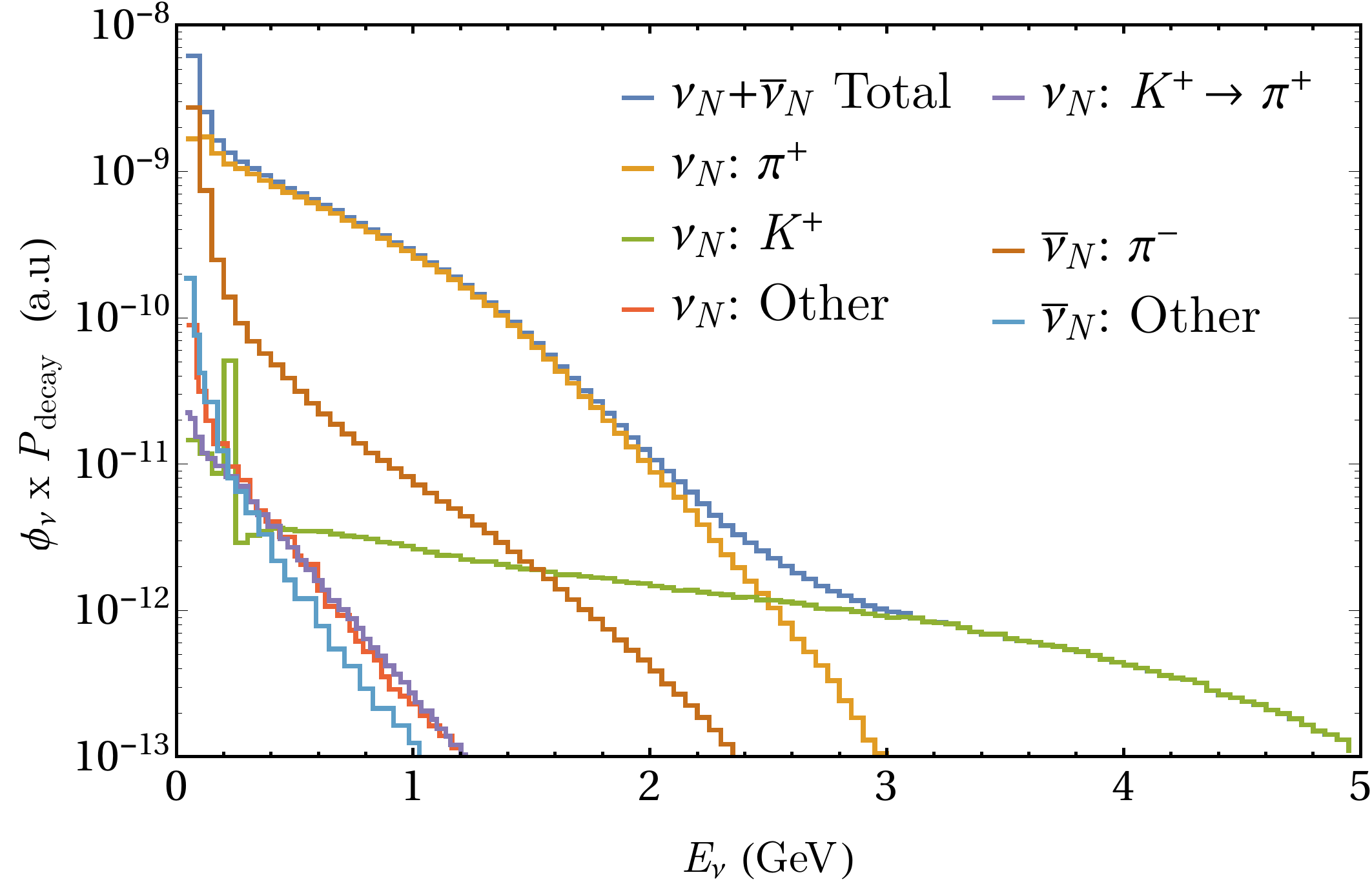}
\end{subfigure}

\caption{\label{fig:flux_plots} Left: The composition of fluxes of $\nu_\mu$
and $\overline{\nu}_\mu$ at \muboone\ with horn in positive polarity (neutrino
mode). ``Other'' refers to contributions primarily from meson decay chains
initiated by meson-nucleus interactions. Right: Fluxes weighted by the
probability to decay inside \muboone, for a sample 25 MeV sterile neutrino with
equal $|U_{e4}|^2 = |U_{\mu 4}|^2$. Requiring that the sterile neutrino decays
inside the detector has the effect of vastly increasing the importance of lower
energy bins, where traditionally cross-section induced background effects are
small.}

\end{figure}

As we are exploring a large parameter space, often this expression takes a
simplified form depending on the size of $\Gamma_\text{T}\lambda/\gamma\beta$:
\begin{align} 
\Gamma_\text{T}\lambda \ll 1\qquad&\qquad P_\text{D} =
e^{-\frac{\Gamma_\text{T}L}{\gamma\beta}}\frac{\Gamma_\text{c}\lambda}{\gamma\beta}
+ \mathcal{O}\left(\Gamma_\text{T}^2\lambda^2\right),\label{eq:prob_dec1}\\
\Gamma_\text{T}\lambda \gg 1\qquad&\qquad P_\text{D} =
e^{-\frac{\Gamma_\text{T}L}{\gamma\beta}}\frac{\Gamma_\text{c}}{\Gamma_\text{T}}
+ \mathcal{O}\left(\frac{1}{\Gamma_\text{T}\lambda}\right),
\label{eq:prob_dec2}
\end{align}
where the rate for slowly decaying particles can be seen to grow with detector
size until a width of $\lambda\sim\gamma\beta\Gamma_\text{T}^{-1}$.
For detectors longer than this scale, the event rate becomes
independent of detector size, as most sterile neutrinos decay within a few
decay lengths.

The spectral flux of sterile neutrinos impinging on a SBN detector,
$\mathrm{d}\phi/\mathrm{d}E$, is estimated as described in \refsec{sec:prod}.
Of crucial importance to this is accurate knowledge of active neutrino fluxes
at all three SBN detectors. These are calculated from published MiniBooNE
fluxes \cite{AguilarArevalo:2008yp}, after scaling by appropriate $1/r^2$
baseline dependence, \eg\ $(470/540)^2 \approx 1.3$ at \muboone. This is
similarly scaled by $1/r^2$ for ICARUS at 600m, however, an additional energy
dependent flux modifier is applied for SBND at 110m to account for the softer
energy spectrum due to the proximity of the detector to the production target
\cite{Antonello:2015lea}. We consider sources of neutrinos that are relevant
including wrong sign neutrinos, smaller sub-dominant $K^+\rightarrow
\pi^+\rightarrow \nu_\alpha$ sources as well as other contributions,
predominantly from meson decay chains initiated by meson-nucleus interactions.
The neutrino spectrum at \muboone\ is shown in the left panel of
\reffig{fig:flux_plots}. 
In the right panel, we also show the effective spectrum of
\emph{decaying} particles at \muboone. As the decay probability
for any given sterile neutrino scales as $1/|P_{\ster}|$, we see an enhancement
of the lowest energy parts of the spectrum. This is in stark
contrast to standard neutrino interaction cross sections, which tend to scale
as $E_\nu$. 
This low-energy bias exaggerates the kinematic differences between our
decay-in-flight signal and the dominant background events.

Finally, the function $W_\text{c}(E)$ is a weighting factor which accounts for
all effects which reduce the number of events in the sample: for example,
analysis cuts or detector performance effects. 
To compute these factors, we run a Monte Carlo simulation of the decays for a
large number of sample events with a given energy. Each sterile neutrino event is
associated with a decay of type $\text{c}$. We then apply experimental analysis
cuts to the decays based on our assumptions about the detector's capabilities
and backgrounds, to produce a spectrum representing the final event sample when
considering events in the bucket timing window (See \refapp{app:bg} for details
of the background analysis). The percentage of accepted events defines the
weight factor for that energy. In this manner the full spectral shape of the signal is
included in the total rate analysis. As a consistency check of our methodology, we
also reproduce in \refapp{sec:ps191} some of the published bounds of PS-191. 

\subsection{\label{sec:backgroundestimate}Background reduction}

In order to estimate the impact of potential backgrounds we have performed a
Monte-Carlo analysis using the neutrino event generator GENIE
\cite{Andreopoulos:2009rq}. This provides generator level information about the
kinematics of the beam-driven backgrounds, with rates normalised off expected
NC and CC inclusive values as published in the SBN proposal. Energy and angular
smearing is then implemented to allow for approximate estimates of the effects
of detector performance to the level necessary for this analysis, without the
need for a full GEANT detector simulation. Energies are smeared according to a
Gaussian distribution around their true MC energies, with a relative variance
$\sigma_E/E = \xi/ \sqrt(E) $, where $\xi$ is a detector dependent resolution.
For this study we take the energy resolution for EM showers, muons and protons
to be 15\%, 6\% and a conservative 15\% respectively, alongside an angular
resolution in LAr of $1.73^{\circ}$ \cite{Antonello:2015lea}. 

\begin{figure}[t]
\center
\includegraphics[width=0.6\textwidth,clip,trim=0 0 0 0]{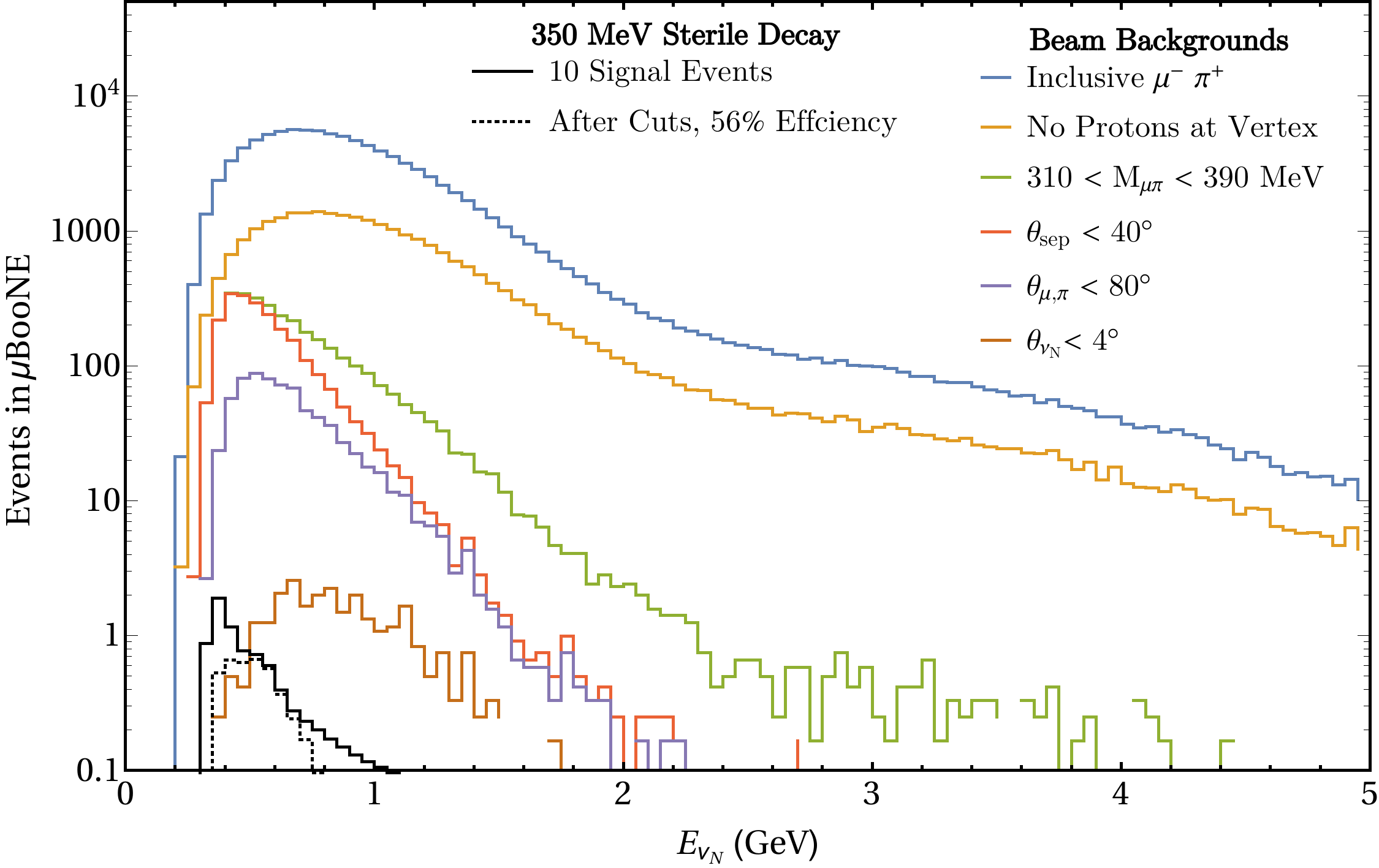}

\caption{\label{fig:mu_pi_cutflow} Reconstructed sterile neutrino energy
spectra for CC$\nu_\mu$ backgrounds in comparison to a 350 MeV decaying sterile
neutrino at \muboone, normalised to 10 signal events. Total expected background
of 98,013 events is reduced to $\approx$ 27 by successive kinematic cuts (as
listed in legend) which utilise the stark differences between decay-in-flight
and scattering kinematics. Further cuts on energy would allow for even greater
reduction. }

\end{figure}

Of utmost importance in all studied channels is the identification of a
scattering vertex, which cleanly indicates that the process is not a
decay-in-flight event. Any hadronic activity localised at the beginning of the
lepton track is a smoking gun for a deep-inelastic or quasi-elastic
beam-related scattering event. Therefore we reject any event containing one or more
reconstructed protons or additional hadrons. For counting
this proton multiplicity we assume a detection threshold of 21 MeV on proton
kinetic energy in liquid Argon \cite{Acciarri:2014gev}, after smearing.
Background events with energies below this threshold and events that do not contain
any protons (such as events originating from coherent pion production) remain a
viable background and further rejection must come from the kinematics of the
final state particles only. The kinematics of such daughter particles
originating from decay-in-flight and backgrounds from scattering events,
however,  have strikingly different behaviour leading to strong suppression
capabilities.

As a representative example of our analysis we discuss the backgrounds
associated with the decay $\ster \rightarrow \mu^\pm \pi^\mp$, the channel with
largest expected beam related backgrounds in all SBN detectors, the dominant
component of which arises from genuine charged current $\pi \mu$ production.
These events can be produced incoherently, often with large hadronic activity
and so will greatly be reduced by the cut on a scattering vertex, or from
coherent scattering, where the neutrino scatters from the whole nucleus.
Coherent cross-sections for these processes have been studied in MiniBooNE
\cite{Wascko:2006tx}, \minerva \cite{Higuera:2014azj} and lately T2K and
cross-sections appear to agree with Monte Carlo calculations based on the
Rein-Sehgal model \cite{Rein:2006di} and generally do not have an additional
hadronic component to cut on. Kinematics of the daughter particle  alone but be
used for background rejection. There has been a noted deficit of forward going
muons \cite{Rein:2006di} in these coherent cross-sections, which is in stark
comparison to the relatively forward behaviour of sterile neutrino decays. 

Furthermore, this channel, and indeed $e^\pm \pi^\mp$, has a powerful
discriminator in the reconstructed invariant mass of the charged particle pair,
\eg\  $M_{l^\pm \pi^\mp}^2=m_l^2+m_{\pi^\pm}^2+ 2(E_l E_\pi -
|P_l||P_\pi|\cos\theta_\text{sep})$ for $\ster\rightarrow \pi^\pm l^\mp$, which
sum to that of the the parent sterile neutrino (within detector resolution),
whereas the background forms a broad spectrum across the energies of the
incoming neutrinos. On top of this strong invariant mass discriminator, these
two body decays allow for reconstruction of the parent sterile neutrino angle
with respect to the beamline which is very close to on-axis, as opposed to the
more isotropic backgrounds. We find that approximately $95$\% of the
reconstructed sterile neutrino angles from these decays are inside a $4^\circ$
cone centred on the beamline. 

We show the effect of our cuts for this channel in \reffig{fig:mu_pi_cutflow},
which ultimately leads to a reduction in the inclusive $\mu \pi$ event rate at
SBND (\muboone and ICARUS) from  1,530,900 (98,013 and 164,716) to  323 (27 and
46) while maintaining a signal efficiency of $56\%$. This level of background
suppression crucially relies on the angular and energy resolution of LAr
detectors, but requires no modification to the current design. 

For a parallel discussion of the backgrounds for the remaining channels see
\refapp{app:bg}.

\subsection{\label{sec:timing}Role of event timing}

\begin{figure}[t] \center \begin{subfigure}[c]{0.68\textwidth}
\includegraphics[width=\textwidth]{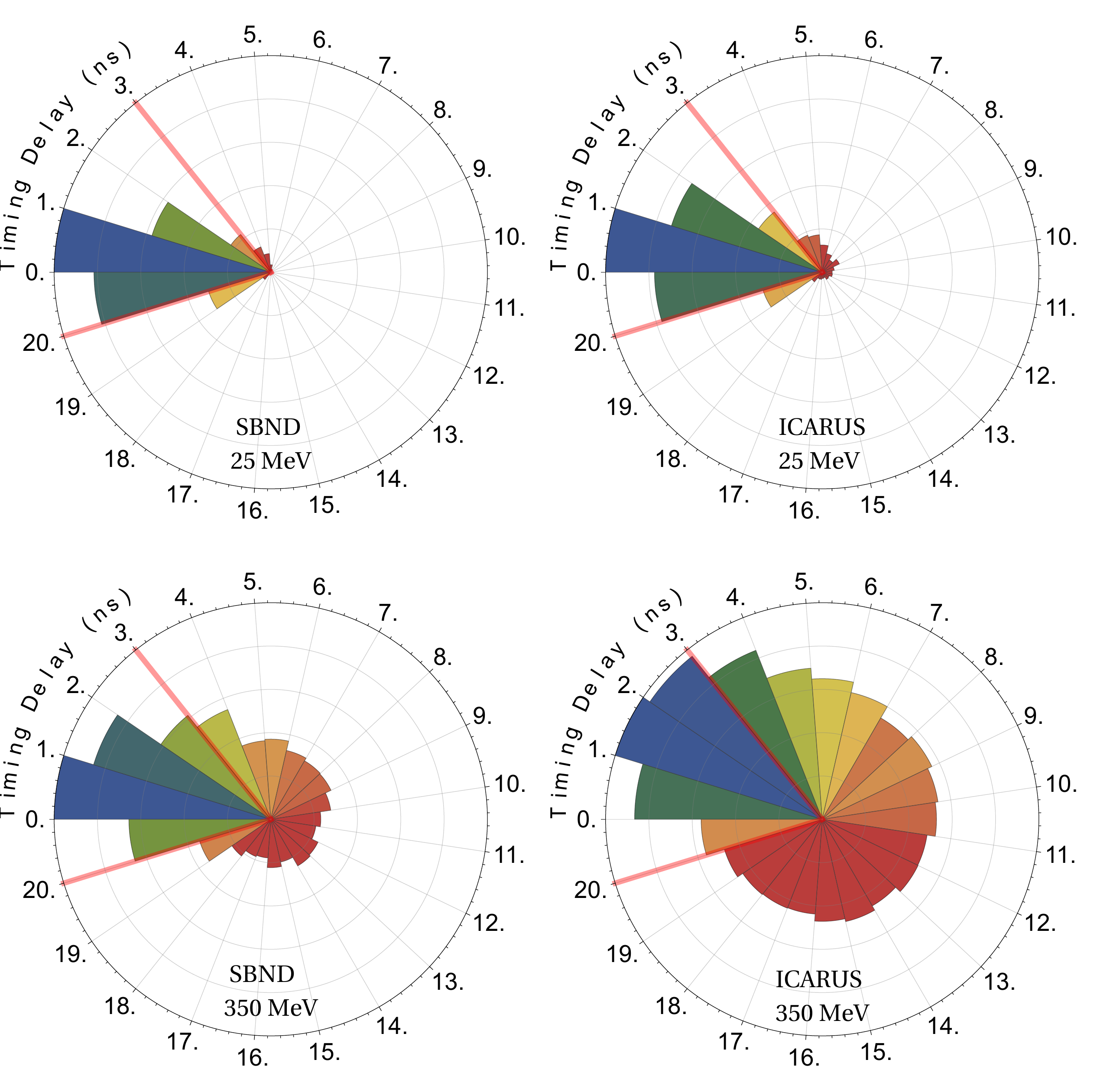} \end{subfigure}
\begin{subfigure}[c]{0.32\textwidth}
\includegraphics[width=\textwidth]{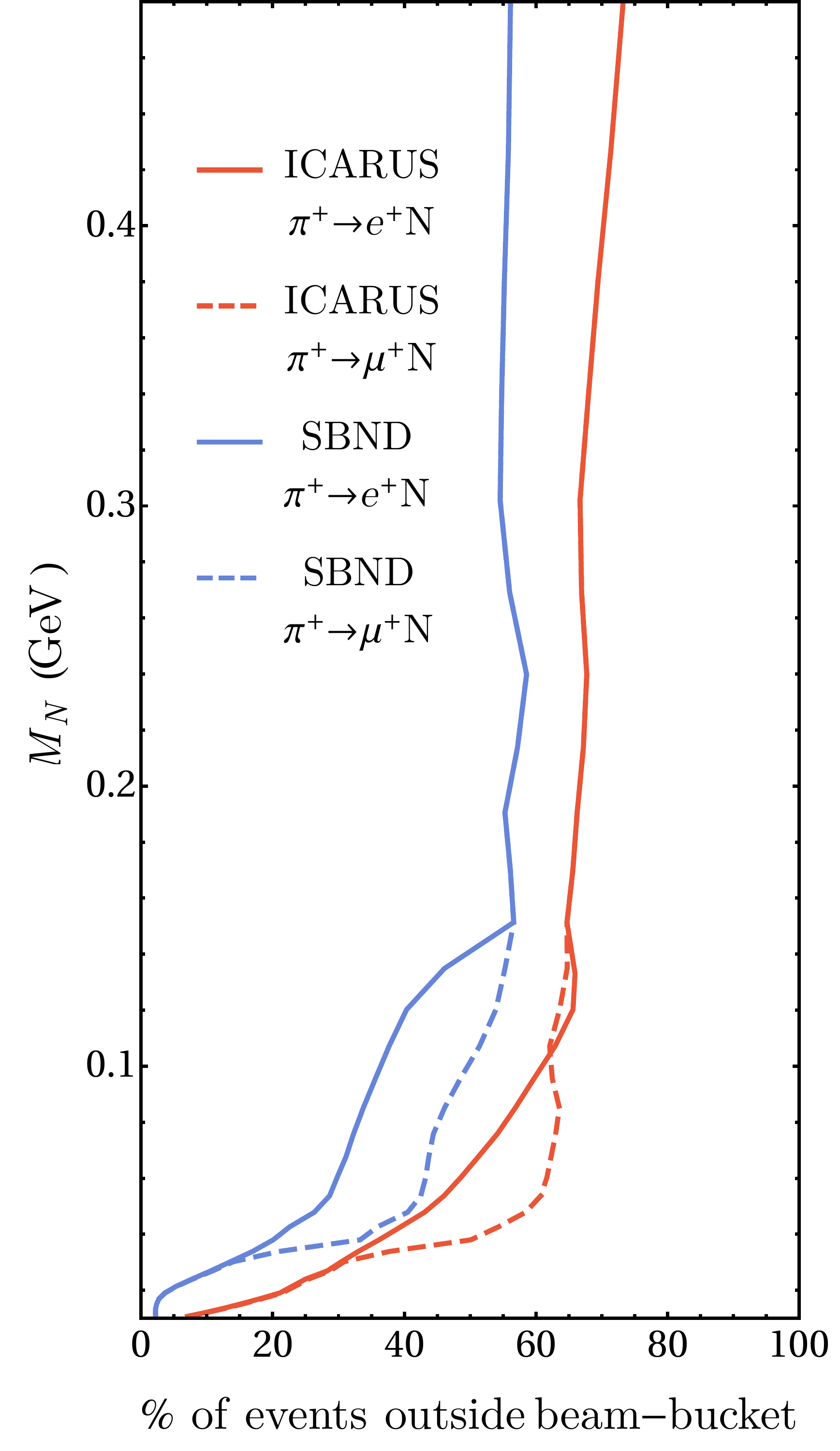} \end{subfigure}

\caption{\label{fig:timing} Left: The timing delay of sterile neutrino decays
in nano-seconds for both a 25 MeV (top) and 350 MeV (bottom) sterile neutrino
at the SBND and and \icarus\ detectors (110 and 600m respectively).  A 4 ns
beam bucket window is shown highlighted in red from 0 to 4 ns, followed by an
additional 17 ns gap. The timings are calculated as a difference to the time of
flight of a active neutrino, assuming the decay occurred in a uniform sample
across the 50m BNB decay pipe. A timing resolution of 1 ns is assumed to smear
the observed events. Right: The percentage of sterile neutrino decay events
that fall into the inter-bucket region as a function of sterile neutrino mass
for SBND and ICARUS, assuming a flux derived from $U_{e4}$ ($U_{\mu 4}$) mixing
in solid (dashed) lines. Both SBND and ICARUS see a sizeable fraction of total
events outside the beam bucket windows when the sterile neutrino mass exceeds
$\approx10$ MeV.  }

\end{figure}

On top of the impressive background rejection capabilities of LAr from
kinematic cuts, there is the potential for an even greater background
suppression by considering the time of arrival of observed events. Although the
drift time of electrons in LAr can be as large as $100$~$\mu$s, the ionisation
and excitation of Argon from the passage of a charged particle also produces
128~nm scintillation light of which there is a nano-second scale contribution
from the decay of the excited state $\text{Ar}^*_2$ \cite{Acciarri:2015hha}.
LAr is transparent to this light, and if the light detection system (LDS)
employed by the SBN detectors has a nanosecond resolution, this can allow for
precise timing to be attached to each TPC triggered event.

Light neutrinos propagate and reach the furthest detector of the SBN complex
after approximately 2 $\mu$s. In the conventional physics program of the SBN,
the timing of these events play an important role in the analysis of
backgrounds, tight timing windows are placed around the 19.2 $\mu$s beam spill
to limit constant rate backgrounds such as cosmogenic events
\cite{Antonello:2015lea}. The LDS of both SBND and ICARUS, however, are
potentially  able to achieve significantly better timing resolution than this,
around $1$--$2$~ns depending on the exact technology used, which potentially
allows for the use of both bucket and spill structure in the background
analysis. The BNB consists of 81 Radio-Frequency buckets of approximately 2~ns
length, separated by 19~ns, to form the 19.2~$\mu$s spill with a frequency of
3~Hz \cite{Antonello:2015lea}.  If this nano-second resolution is indeed
achieved, it allows for events in individual buckets to be identified.  Such a
nano-second resolution was achieved previously by the PMTs used in MiniBooNE
\cite{Antonello:2015lea}, with potential for improvement in the next generation
SBN detectors. \muboone\ is omitted from considerations of timing as its
achievable timing resolution is lower at around $10$~ns \cite{Katori:2013wqa}.

As particles with finite rest mass, heavy sterile neutrinos will propagate at
subluminal speeds which can produce observable timing delays.  This effect
begins to become relevant when the sterile neutrinos have MeV-scale masses and
above. As the flux of decaying sterile neutrinos is inversely proportional to
its momentum after convolving with their decay probability, many of these low
energy sterile neutrinos are travelling at sufficiently slower speeds than
their light counterparts to be distinguishable. Shown on the right of
\reffig{fig:timing} is the fraction of events that are expected to fall outside
the  bucket window in both SBND and ICARUS. For the purposes of this study we
define the beam-correlated window to be a $6$~ns period, $2$~ns each side of
the $2$~ns beam bucket. The exact width of the beam bucket window can be
modified if studying channels with low expected backgrounds.
In this section, we consider only the timing of events relative to the bucket
window\footnote{Absolute arrival times could in principle be used, but this
would require good synchronisation between geographically separated clocks.
Alternatively, the relative timing between signal and beam-related backgrounds
could be used. However, we do not consider these options further.}, a structure
which repeats every 21 ns. Delayed events can be observed in any subsequent
window, producing a 21-fold degeneracy in their reconstructed arrival time.
This lends a cyclical nature to the timing information, with a distinctive
structure at the different detector sites for larger masses. Some illustrative
timing distributions are shown on the left of \reffig{fig:timing} for a $25$
and $350$ MeV sterile neutrino.

We find a significant proportion of sterile neutrino events distributed
throughout the inter-bucket region. Events which fall into the beam-bucket
timing window have to be analysed on top of all known beam-related backgrounds,
but events in the inter-bucket window have significantly reduced
beam-correlated backgrounds. 
For larger masses, we have shown that the majority of events fall into these
regions, and this may allow for a low background search strategy for decaying
sterile neutrinos. Instead of beam-correlated backgrounds, the constant rate
backgrounds will limit the sensitivity for this analysis. Understanding these
backgrounds in detail is beyond the scope of this work; however, we expect the
strongly forward kinematics, combined with \emph{in situ} beam-off measurements
will allow for a very low backgrounds to be obtained.

In the following sections, timing information will inform our work in three
ways. First we will compute SBN's sensitivity to decaying sterile neutrinos
assuming the full backgrounds, reduced only by the cut-based analysis discussed
previously.  This is a proven sensitivity, applicable for all sterile neutrino
masses and detectors and is independent of the attainable timing resolution.
Secondly, we compute a backgroundless sensitivity. This can be seen as either
the result of improved analysis techniques, or as the inclusion of timing
information at SBND and ICARUS for the largest masses. Finally, in
\refsec{sec:timing}, we will study the use of the timing information itself to
constrain the underlying model of decaying sterile neutrinos.

\subsection{\label{sec:eventspectra}Event spectra}

\begin{figure}[t]
\centering
\includegraphics[width=0.45\textwidth]{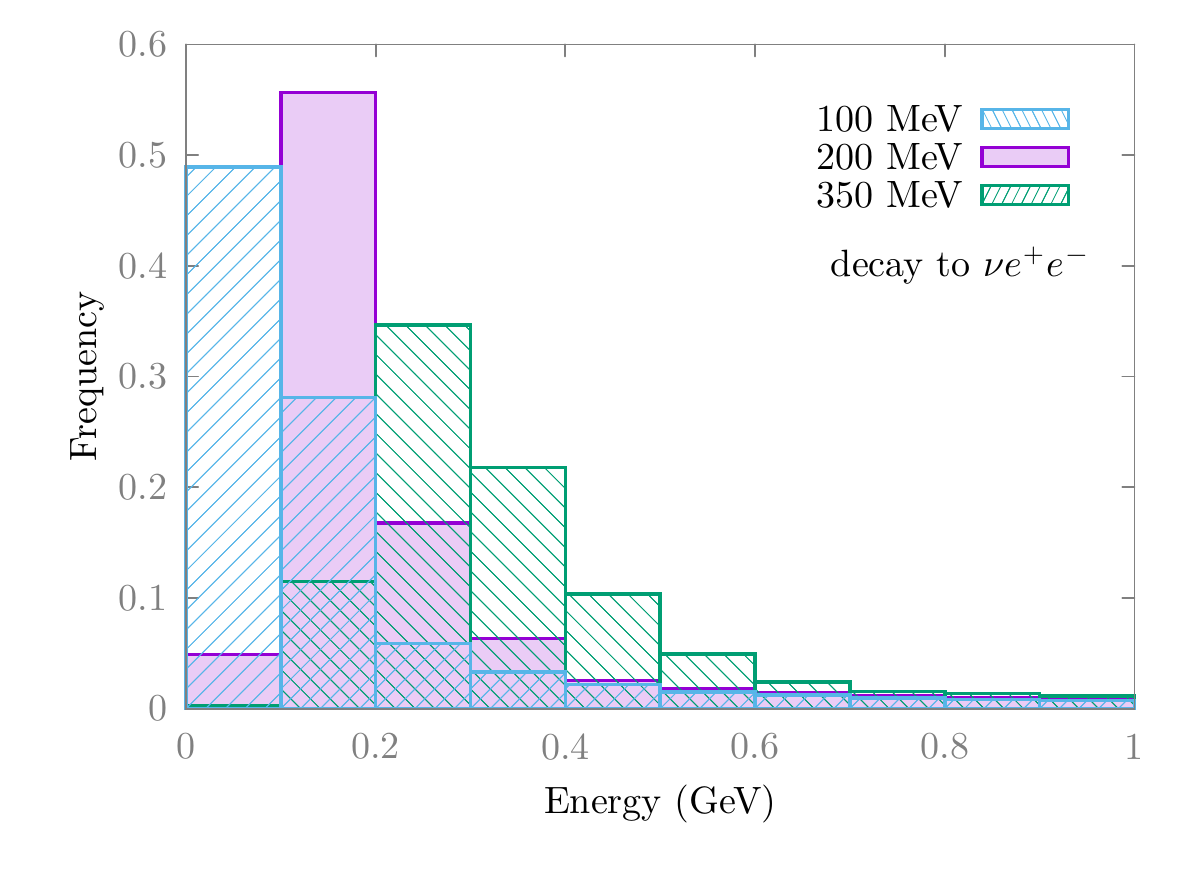}
\includegraphics[width=0.45\textwidth]{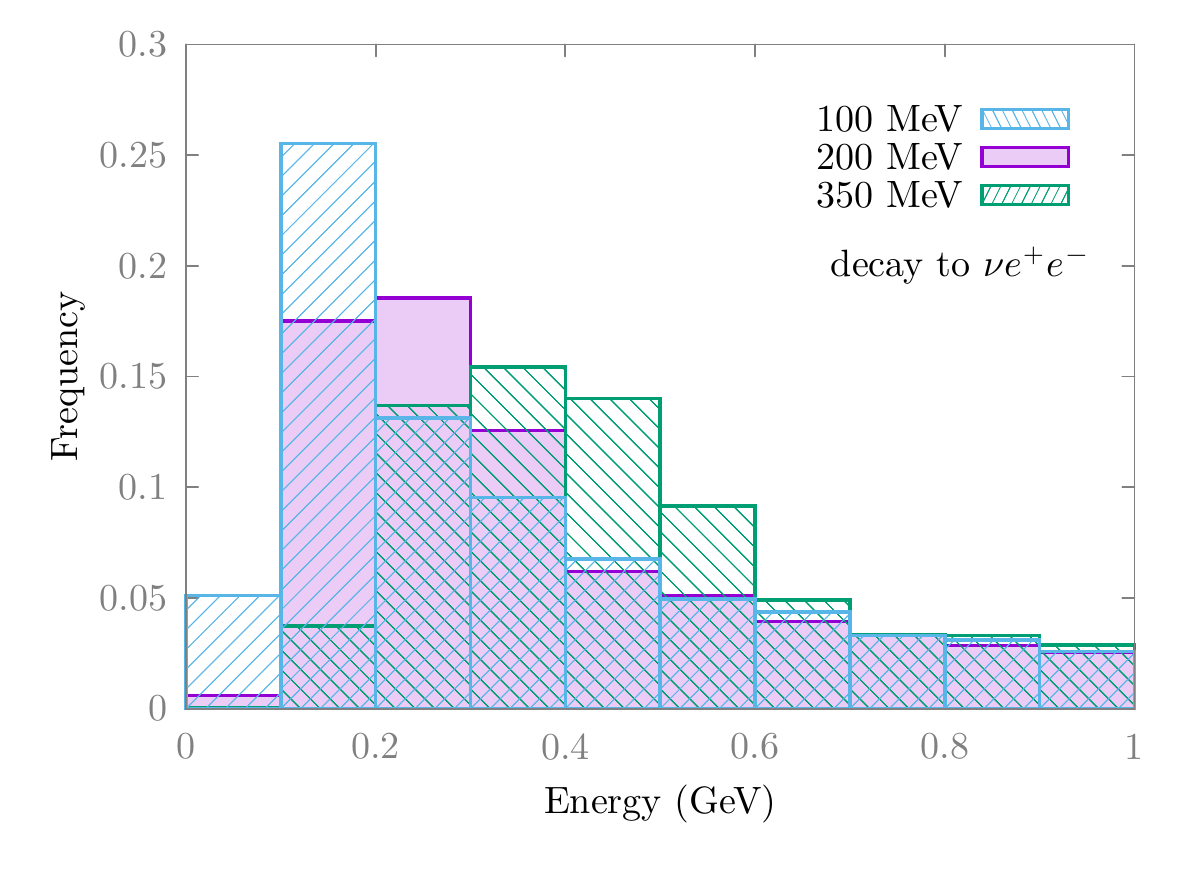}\\

\includegraphics[width=0.6\textwidth]{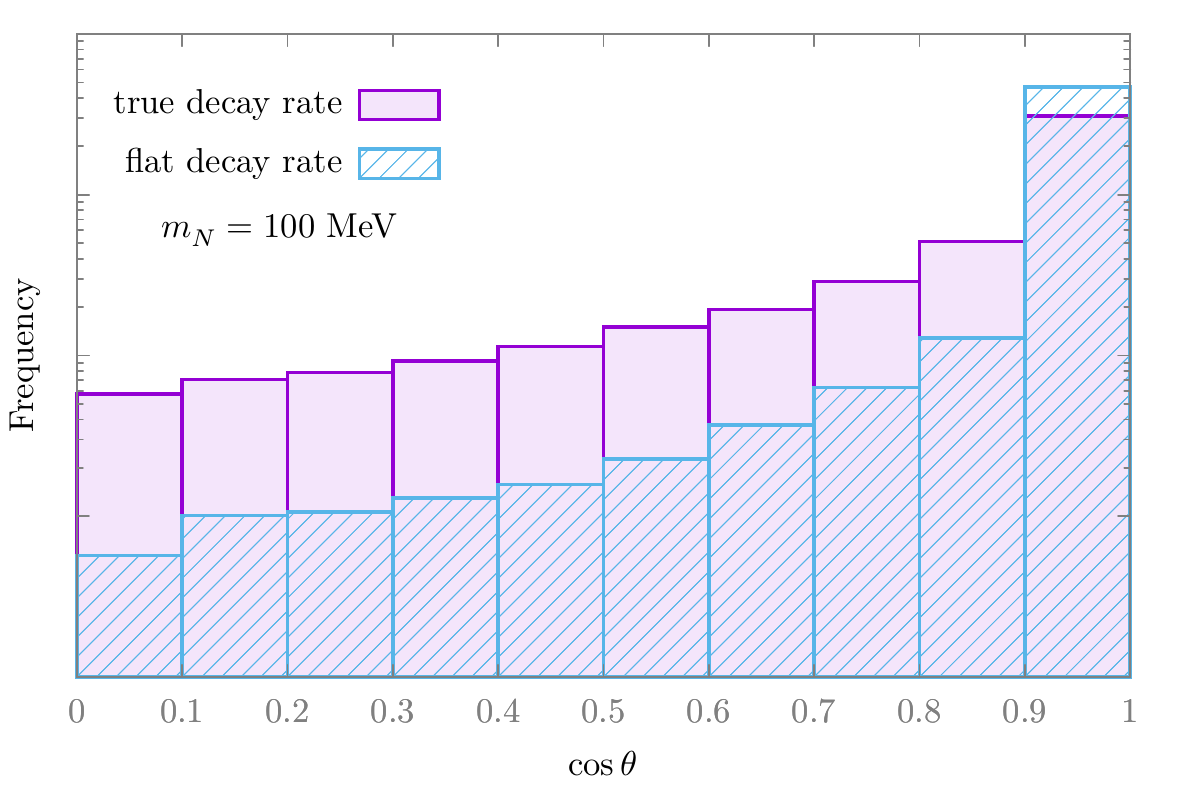}

	\caption{\label{fig:spectrum_ee} Top: Characteristic spectra for the total energy of observed  $e^+e^-$ pairs seen at \muboone\ produced in the $\ster \to \nu e^+e^-$ decay mode, for three representative masses. In the left panel, the spectra have no analysis cuts or detector reconstruction effects applied, while on the right these are included, reducing the number of lowest-energy events. Bottom:  The expected angular distributions for the $e^+e^-$ pair from a sterile neutrino of mass $m_\ster = 100$ MeV. The red histogram shows the true expected distribution, while the blue histogram shows the distribution if we do not take into account the preferential
		decay rate for lower energy particles, instead using an energy independent
		decay rate.}
\end{figure}

\begin{figure}[t]
	\center
		\includegraphics[width=0.5\textwidth]{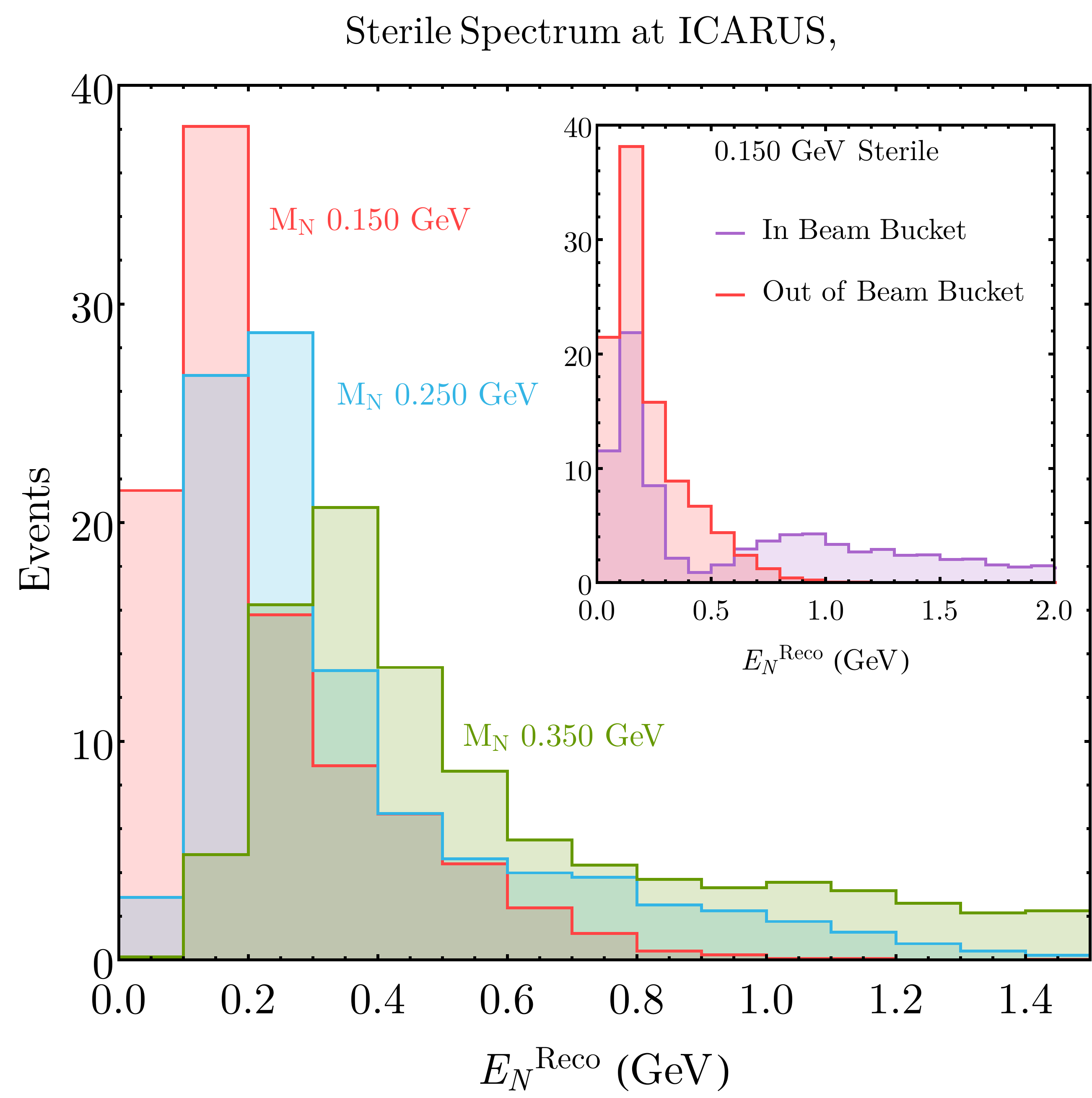}
	\caption{\label{fig:spectrum_epi} 
		Characteristic spectra for the reconstructed neutrino energy for $\ster\rightarrow e^\pm \pi^\mp$ and a sterile neutrino masses of 150, 250 and 350 MeV. The insert shows the stark differences in spectrum when one considers events falling within the beam bucket window and without.}
\end{figure}

The differential distributions from heavy sterile neutrino decay tend to
produce distinctive low-energy distributions of events with an appreciably
forward direction. 
The tendency towards low energies is predominantly due to the higher decay
rates of low-energy particles, which leads to factors of $1/E_\nu$ in the event
rate formula \refeq{eq:prob_dec1}.
The forward trajectory is inherited from the kinematics of a boosted object
decaying in flight. However, this effect is slightly mitigated by the
preference for lower energy decays, meaning that lower energy sterile neutrinos
are more likely to decay, which are the least boosted objects.

We show an example of a distribution for electron-positron production in the
top left panel of \reffig{fig:spectrum_ee}. For the lowest masses that we
consider, almost all events have energies below $0.5$ GeV, in this case
illustrated by the blue histogram. The distribution tends towards larger
energies as the mass of the sterile neutrino increases, but for sterile
neutrino masses less than the kaon mass, never produces significant numbers of
events above $1$ GeV. As we can see in \reffig{fig:spectrum_ee}, the number of
events in the lowest energy bin is strongly indicative of the mass of the
parent particle, and therefore the lowest energy events will play a strong role
in model discrimination. However, in the cut-based analysis which we outline in
\refapp{app:bg} the lowest energy event distribution is significantly reduced
due to poor efficiency's at low-energy in our cuts, as can be seen in the top
right panel of \reffig{fig:spectrum_ee}. In \reffig{fig:spectrum_epi} we show
the same distributions for electron-pion production, noting similar spectral
features of the electron-positron channel. Indeed this behaviour qualitatively
exists in all channels studied. In the inset of \reffig{fig:spectrum_epi}, we
highlight the differences that an accurate timing resolution can give, with the
in-bucket and out of bucket spectra showing very significant differences.
Through optimisation of this part of the analysis, we expect the sensitivity to
these models can be improved; however, this is beyond the scope of the present
work.  The angular spectrum is expected to be very informative in these models,
and the events are predict to align with the beam direction. The red histogram
in the lower panel of \reffig{fig:spectrum_ee} illustrates an expected
distribution for the four momentum of a $e^+e^-$ pair in the decay $N\to\nu
e^+e^-$. We compare it to the expected distribution found for events without
the low-energy biasing effect of decay-in-flight, with an unphysical energy
independent decay rate (denoted `flat', shown in blue). Not only does the
decay-in-flight probability lead to a lower energy events, but it also makes
the angular distribution less forward.

\section{\label{sec:sensitivities}Results}

In this section, we present the results of our simulation for two
analyses. In the first, we compute exclusion contours which could be expected
to be set by SBN if no signal is seen. We compute these for all decay modes
presented in \reffig{fig:branchingratios}. Our second analysis considers the
phenomenological potential of energy and timing spectral information at the SBN
experiment if a potential signal is observed. 

Due to its proximity to the BNB target, SBND provides the majority of the
statistics, and hence the sensitivity, to sterile neutrino decays. The addition
of \muboone\ and ICARUS increases the event rate by approximately 6\%. However,
the power of the three detector SBN setup arises not from the increased
statistics, but rather from the additional phenomenology of a multi-baseline
experiment. We show below that ICARUS, being the furthest detector, can play an
important role in study of any observed signals through precision timing
measurements.  Similarly, although \muboone\ contributes a small fraction to
the raw number of sterile neutrino events expected, the \muboone\ experiment is
significantly more advanced than its two SBN counterparts. At the time of
writing, \muboone\ has already collected close to $50$\% of its planned POT
(around $3.5\times 10^{20}$ POT) and has already presented its first results on
$\nu_\mu$ CC inclusive and $\nu_\mu$ CC$\pi^0$ interactions
\cite{mubooneneutrino}. As such, \muboone\ is in a unique position in that it
has the potential to observe any excess in advance of SBND or ICARUS and thus
to inform a possible search using the full SBN programme.
Non-observation of any excess at \muboone\ would not negatively effect the
subsequent search for new physics at SBND or ICARUS significantly, however, as
a large fraction of the testable parameter space is only accessible through the
enhanced exposures of the full SBN programme.

\begin{figure}[t]
\center
\includegraphics[width=1\textwidth]{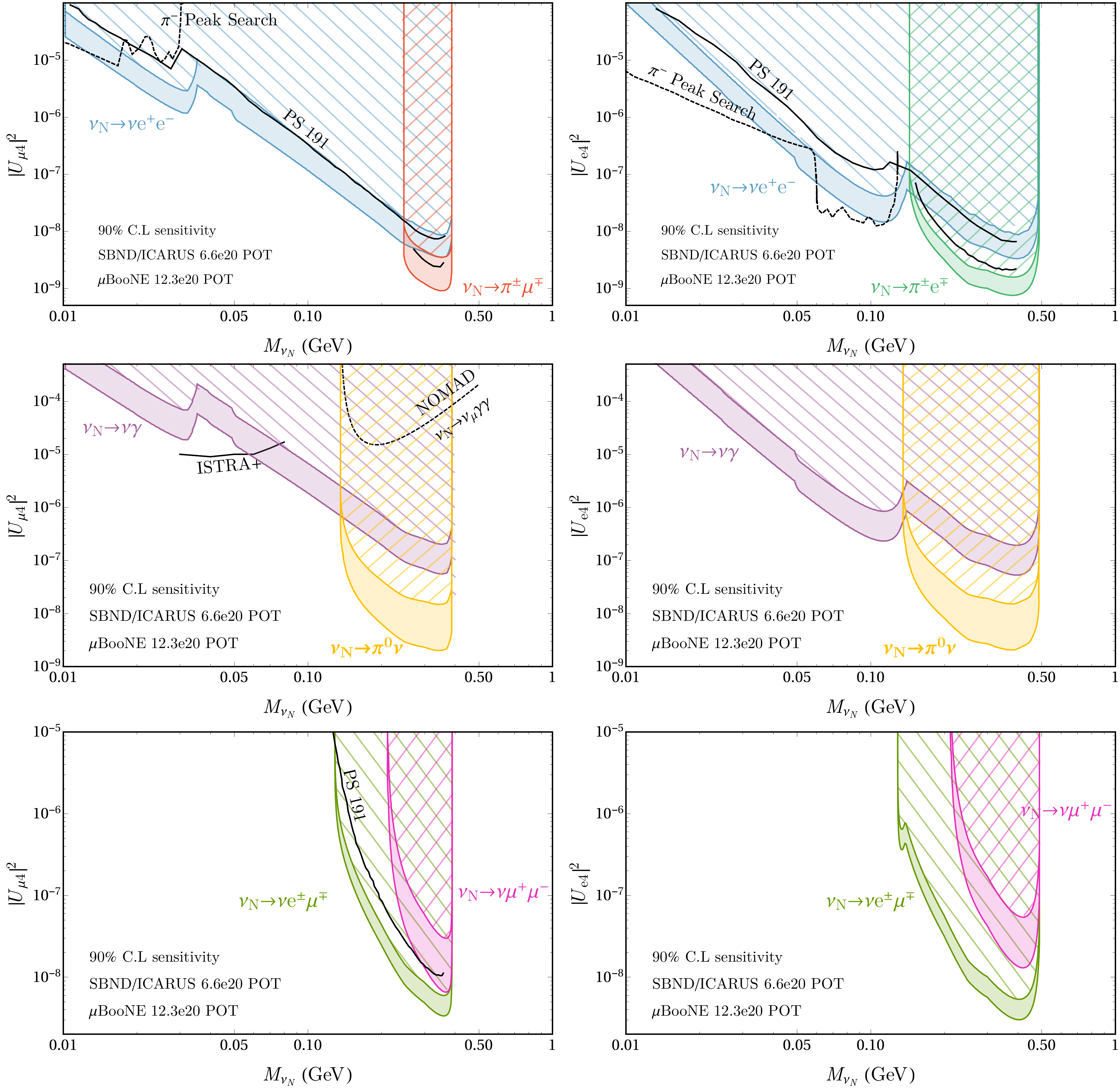}

\caption{\label{fig:band_sbn}The predicted  90\% CL  upper limit contours for
the combined SBN detectors. Shown also in black solid lines is the prior best
bounds from PS-191, scaled to show bounds on the minimal extension as discussed
here, as well as bounds from lepton peak searches in pion and kaon decay
\cite{PhysRevD.46.R885,PhysRevLett.68.3000} (dashed black lines). Note that the
peak searches are only valid when bounding pure mixing combinations, \eg\
$|U_{\mu 4}|^2$ and not $|U_{\mu 4}||U_{\tau 4}|$. The photonic channels have
little or no direct bounds, with ISTRA+ bounding the radiative decay
\cite{Duk:2011yv} and reinterpreted $\ster \rightarrow \nu \gamma \gamma$
bounds at NOMAD on $\ster \rightarrow \nu \pi^0$ \cite{Gninenko:1998nn}. In all
panels, the mixing matrix elements not shown on the $y$-axis are zero.}

\end{figure}

\subsection{Limits on sterile neutrino mixing}
We have computed the bounds that SBN could place on sterile neutrino
mixing-matrix elements for all kinematically accessible visible decays.
\reffig{fig:band_sbn} presents the results of our analysis assuming a combined
$6.6\times 10^{20}$ POT at SBND and ICARUS, and $13.2\times10^{20}$ POT at
\muboone.  We plot the predicted upper limits on sterile neutrino mixing for
the leptonic decay channels $\ster \rightarrow \nu_\alpha e^+e^-$, $\ster
\rightarrow \nu_\alpha \mu^+ \mu^- $ and $\ster \rightarrow \nu_\alpha \mu^+
\mu^- $ as well as the semi-leptonic and photonic channels $\ster \rightarrow
l^\mp \pi^\pm$,  $\ster \rightarrow \nu_\alpha \pi^0$ and $\ster \rightarrow
\nu_\alpha \gamma$. The plot on the right (left) assumes that the mixing-matrix
element with the electron (muon) flavour is dominant. The $y$-axis refers to a
single mixing element, $|U_{\alpha 4}|^2$, but each bound is equally applicable
to the combination $|U_{\alpha 4}| |U_{\tau 4}|$, as although production must
proceed through electron-neutrino or muon-neutrino mixing, the decay can take
place through $U_{\tau 4}$ driven processes.  The lower solid coloured lines
are the background-less 90\% CL upper limit contours defined as 2.44 events
following the procedure of \refref{Feldman:1997qc}.  This represents the best
expected sensitivity at the SBN program, assuming perfect signal efficiency.
We also present the results of the cut based background analysis discussed in
\refsec{sec:backgroundestimate} (upper solid coloured lines).  
Depending on the optimisation of the analysis, including the possibility of
using improved timing information, we expect the ultimate sensitivity to be
within the solid-shaded region, lying between the proven cut-based sensitivity
and the backgroundless one. 

The increased event rates at SBN compared to those of PS-191 allows for an
improvement of the bounds on $|U_{e4}|^2$ and $|U_{\mu 4}|^2$ in all channels
studied over wide regions of parameter space. The strongest bounds come from
the semi-leptonic $\ster \rightarrow l^\pm \pi^\mp$ searches, where
mixing-matrix elements greater than $|U_{e4}|^2 \leq 10^{-9}$ can be excluded
at the 90\% CL for $m_\ster \approx 0.350$ GeV. The bounds have the potential
to improve upon the $\pi^-$ peak search bounds for $m_\ster \leq 0.033$ GeV and
$m_\ster \leq 0.138$ GeV for muon and electron mixing respectively, if the
backgrounds can be further suppressed, possibly through the use of timing
information.

Additionally, we show that the previously poorly bounded photonic-like channels
$\ster \rightarrow \nu_\alpha \pi^0$ and $\ster \rightarrow \nu_\alpha \gamma$
can be probed across the entire parameter space, providing new constraints on
exotic sterile neutrino signatures. The potential beam-related backgrounds are
large for these photonic channels, the effect of which is a much wider
separation between our cut-based limits and the optimal ones. These photonic
channels allow SBN to probe the electromagnetic nature of the sterile neutrino,
placing bounds on any models containing enhanced couplings to photons. 
For sterile neutrinos whose mass lies $m_\pi^0 \leq m_\ster \leq m_\pi^\pm
+m_\mu$ and mix primarily with muons, the $\ster \rightarrow \nu_\mu \pi^0$
channel can extend the limits beyond that of the traditional $e^+e^-$ searches
to probe new parameter space, even in the purely minimal model. For sizeable
$U_{e4}$, the $\pi^0$ bounds are less powerful than that of the semi-leptonic
$\ster \rightarrow e^\pm \pi^\mp$ when one assumes the minimal model.

\begin{figure}[t]
\center
\includegraphics[width=1\textwidth]{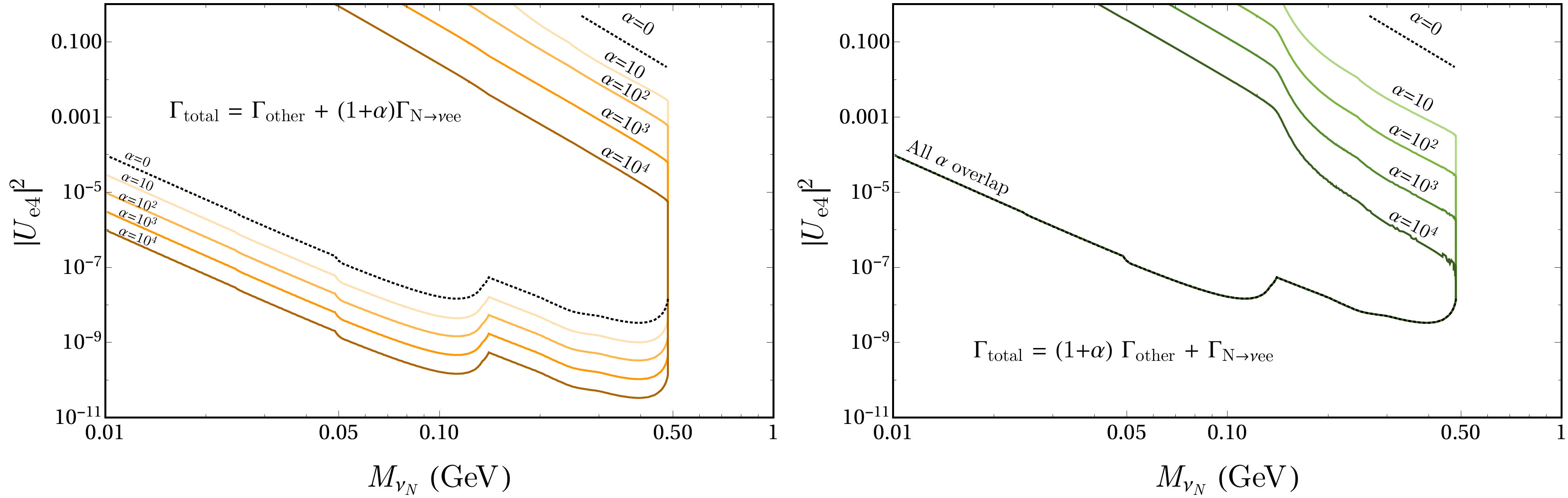}

\caption{\label{fig:sbn_scale}The 90\% CL contours for the decay
$\ster \rightarrow \nu e^+ e^-$ assuming dominant $U_{e4}$ mixing in
non-minimal scenarios, at leading order in $\lambda/L$. On the left, we enhance
the rate of the $e^+e^-$ channel itself by factors shown in the labels while
keeping all other decay rates constant. The excluded region remains roughly
constant but shifts downwards. On the right, we show the effects of keeping the
$\ster \rightarrow \nu e^+ e^-$ decay rate constant, but enhancing the total
decay rate. The sensitive region shrinks quickly as $\alpha$ increases,
allowing non-minimal models to escape detection.}

\end{figure}

Although we have plotted the limits on mixing angles in \reffig{fig:band_sbn}
in terms of the parameters of the minimal model, they are model independent in
the sense that an enhanced decay rate in that channel would only alter the
interpretation of the $y$-axis. If the enhancements to the decay rates are
modest, to reinterpret any bounds on \reffig{fig:band_sbn} in the context of a
non-minimal extension in which the channel of interest is enhanced by
$(1+\alpha)$ then the quantity bounded on each vertical axis is given
approximately by  $|U_{\alpha 4}|^2/\sqrt{1+\alpha}$ as discussed in
\refsec{sec:bounds}. 
However, for larger enhancements, the lower-bound on the
mixing-matrix element must also be considered. In \reffig{fig:band_sbn},
this bound lies at large values of $|U|^2$, and is not shown in
the plots, but it is also affected by an enhanced decay rate and can become
relevant of reasonable enhancements. This can be seen in the left panel of
\reffig{fig:sbn_scale}, where we show the region of parameter
space that SBN could exclude when studying the decay mode $\ster \rightarrow
\nu e^+ e^-$ as we increase its decay rate by factors of $10,10^2,10^3$ and
$10^4$. As was shown analytically in \refsec{sec:bounds}, the upper bound
scales as $1/\sqrt{1+\alpha}$ as the number of events in the
detector increases. However, the enhancement eventually leads to significant
beam attenuation before the detector. This alters the lower bound, which begins
to move to smaller values of the mixing-matrix element, opening up a region of
parameter space in the top-right of the plot. In the right panel of
\reffig{fig:sbn_scale}, we show an alternative non-minimal model in which the decay rate 
$\Gamma\left(\ster \rightarrow \nu e^+ e^-\right)$ is held constant, but the
total decay rate is enhanced. This could be due to the enhancement
of a decay to visible  or invisible final states. In this scenario, the upper
bound remains unchanged as the rate is enhanced (to leading order in
$\lambda/L$), but the enhanced total decay rate leads to beam attenuation and
fewer sterile neutrinos reach the detector.  Eventually, the lower bound is reduced
significantly, and the experiment loses sensitivity over much of the parameter
space of the minimal-model. Although an enhanced total decay rate could produce
a larger visible signal in another channel, or indeed in another experiment, if
the decay is predominantly to three neutrinos or dark sector particles many
existing bounds may not apply.  We note that enhancements on the scale of
$\alpha=10^4$ could be expected if the novel decay proceeds without mixing
suppression. Every bound presented in \reffig{fig:sbn_scale} can be
reinterpreted in terms of these non-minimal models using the scalings as discussed in \refsec{sec:bounds}, and highlights why searching
across the whole parameter space is necessary in all kinematically allowed
decay channels.
	
\subsection{\label{sec:timing_physics}Timing information to study an observed signal}

In addition to being able to reduce beam-related
backgrounds, a precise knowledge of the timing of any observed events can also
be used to discriminate between potential models and aid parameter estimation.
If a potential signal is observed, it would be highly desirable to
establish whether the excess is associated with a heavy particle travelling
from source to detector. An analysis based on the energy spectrum alone would
struggle with this determination --- we could not discount mis-understood
beam-related backgrounds, unknown nuclear effects, or other models that mimic
the low-energy spectrum. The angular distribution of events would also be highly
informative, we have seen that heavy particle decays are likely to be
associated with collimated decay products, but this would be only indirect
evidence of a heavy particle, and could be associated with other models. For
example, active neutrino scattering via a light mediator could also mimic this
behaviour.
However, as all beam-related backgrounds will be correlated with the Booster
proton buckets, the observation of events with times outside of the BNB beam
bucket window (and travelling in a forward direction) would be a smoking gun
signal of a sub-luminal propagating parent. 

We estimate the required timing
resolution by simulating the distribution of arrival times for a given sterile
neutrino mass.  We then compute the compatibility of this data with a
beam-bucket hypothesis, where all event timing is consistent (within errors)
with being within the beam-bucket. We only study the shapes of these timing
distributions, allowing the normalization to float, and in this sense the
beam-bucket hypothesis encompasses all sources of particles which would appear
beam-correlated. The beam-bucket hypothesis is defined as the assumption that
all events originate in a 6 ns window surrounding the BNB beam spill, smeared
by a Gaussian with a width of the assumed time resolution. We define our test
statistic  as \cite{Agashe:2014kda} 
\[ t_m = -2 \ln \left(\mathcal{L}\right) = 2 \sum_{i=1}^N \left\{ \mu_i(m)-n_i
+n_i \ln\left[\frac{n_i}{\mu_i(m)}\right ]  \right\}, \]
where $\mu_i(m)$ is the expected number of events in bin $i$ if the sterile
neutrino is of mass $m$. Using this statistic we have run a binned Maximum
Likelihood analysis of the reconstructed time of arrival $\Delta T$, assuming
events are Poisson distributed. 
We compute the distribution for $t_m$ by Monte Carlo to ensure
that we account for all non-gaussianity in the likelihood function.

\begin{figure}[t]
\center
\begin{subfigure}[t]{0.5\textwidth}
\includegraphics[width=\textwidth]{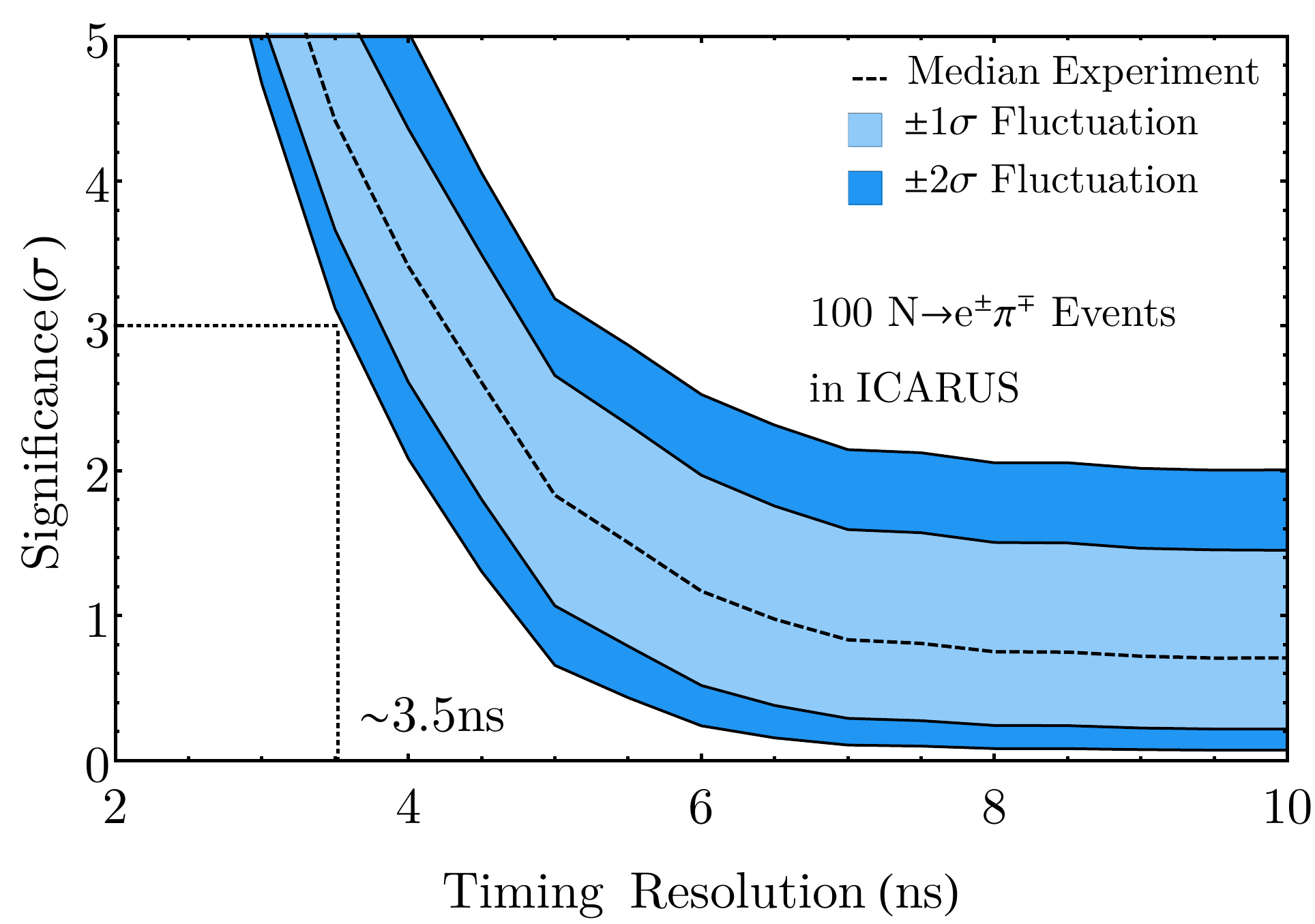}
\end{subfigure}%
~
\begin{subfigure}[t]{0.5\textwidth}
\includegraphics[width=\textwidth]{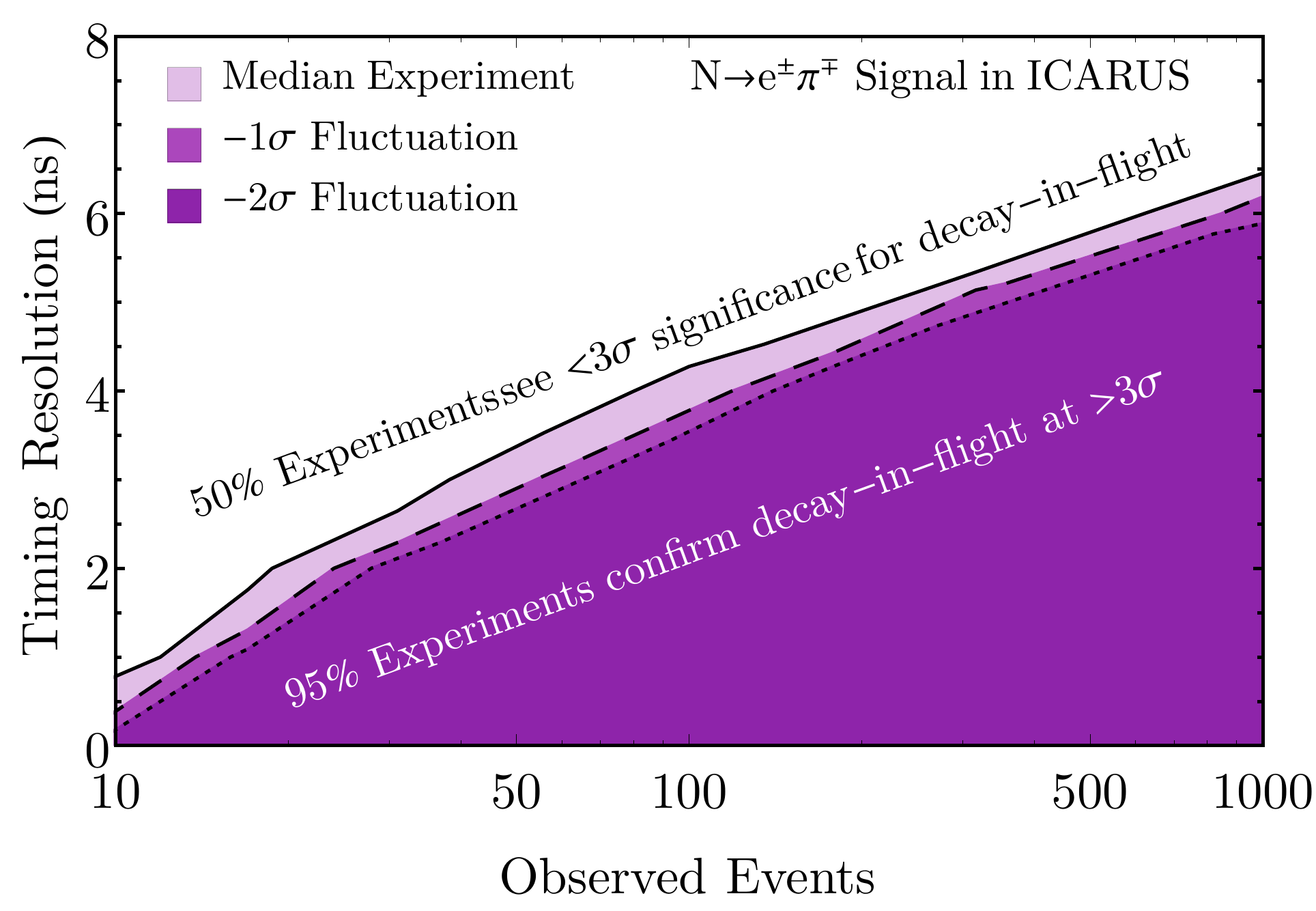}
\end{subfigure}

\caption{\label{fig:hockey}Left: Expected significance at which ICARUS can
exclude a beam-correlated origin from timing information alone, as a function
of assumed timing resolution. This assumes a hypothetical signal of 100 $e^\pm
\pi^\mp$ events consistent with $\ster \rightarrow e^\pm \pi^\mp$.  Right: As
the number of observed events goes down, it is significantly harder to
establish a time delay. We show the timing resolution required for a given
number of events for the median experiment (solid line) as well as for
$1\sigma$ (dashed line) and 2$\sigma$ (dotted line) downward fluctuations.}

\end{figure}

As the timing is solely a
function of the initial sterile neutrino energy and mass, these results hold
for all channels studied. Without loss of generality, we restrict
our discussion to the semi-leptonic channel $\ster\to e^\pm \pi^\mp$. On the
left panel of \reffig{fig:hockey}, we show the timing resolution
required to exclude the beam-bucket hypothesis at a given statistical
significance. This plot assumes that ICARUS has observed an excess of 100
events due to a $300$ MeV sterile neutrino. To guarantee that ICARUS can reject
the beam-bucket hypothesis at at least $3\sigma$ significance in 95\% of
pseudo-experiments, we require a timing resolution of $\leq$ $3.5$ ns.
As the number of observed events increases, the timing resolution
required to rule out a beam-correlated origin decreases, as we show in the
right panel of \reffig{fig:hockey}.

Although establishing that a signal arrived outside of the beam window would be
an exciting sign of new physics, it would not necessarily establish a heavy
propagating parent.
For example, if an unaccounted for process had a fixed time delay with respect
to the neutrino beam, $\Delta_t$, such as the relaxation time of an excited
atom, it could produce events in the inter-bucket region for $\Delta_t \approx
\mathcal{O}$(ns). Similarly, other exotic BSM physics could be the source a
fixed time decay signature without relying on a heavy propagating sterile
neutrino. The scenario described in \refref{Gninenko:2009ks,Gninenko:2010pr} is
one such case, it considers a sterile neutrino produced inside the detector
through neutral current scattering of an active neutrino. The heavy particle
promptly decays, with a decay length of the order $\mathcal{O}$(1)~m, producing
the visible signal.  Although the sterile neutrinos are produced inside the
detector with no timing delay from active-neutrino scattering, the finite
lifetime of these particles could lead to events falling in the inter-bucket
window. 

However, in both the BSM scenarios as well as generic backgrounds with a fixed
timing delay, the temporal spectra of event arrival time would be expected to be
constant across all three detectors.  The SBN program is perfectly designed to
account for this, however, through its multiple detectors at different
baselines, as if the excess is indeed due to heavy particle propagation, then
the sterile neutrino would have to travel further to reach each subsequent
detector. This leads to observable shifts in the arrival timing spectra at each
experiment. In particular ICARUS, would be most suited to studying heavy
particle propagation, as particles must travel approximately 6 times further
before detection.

\begin{figure}[t]
\center
\begin{subfigure}[t]{0.5\textwidth}
\includegraphics[width=\textwidth]{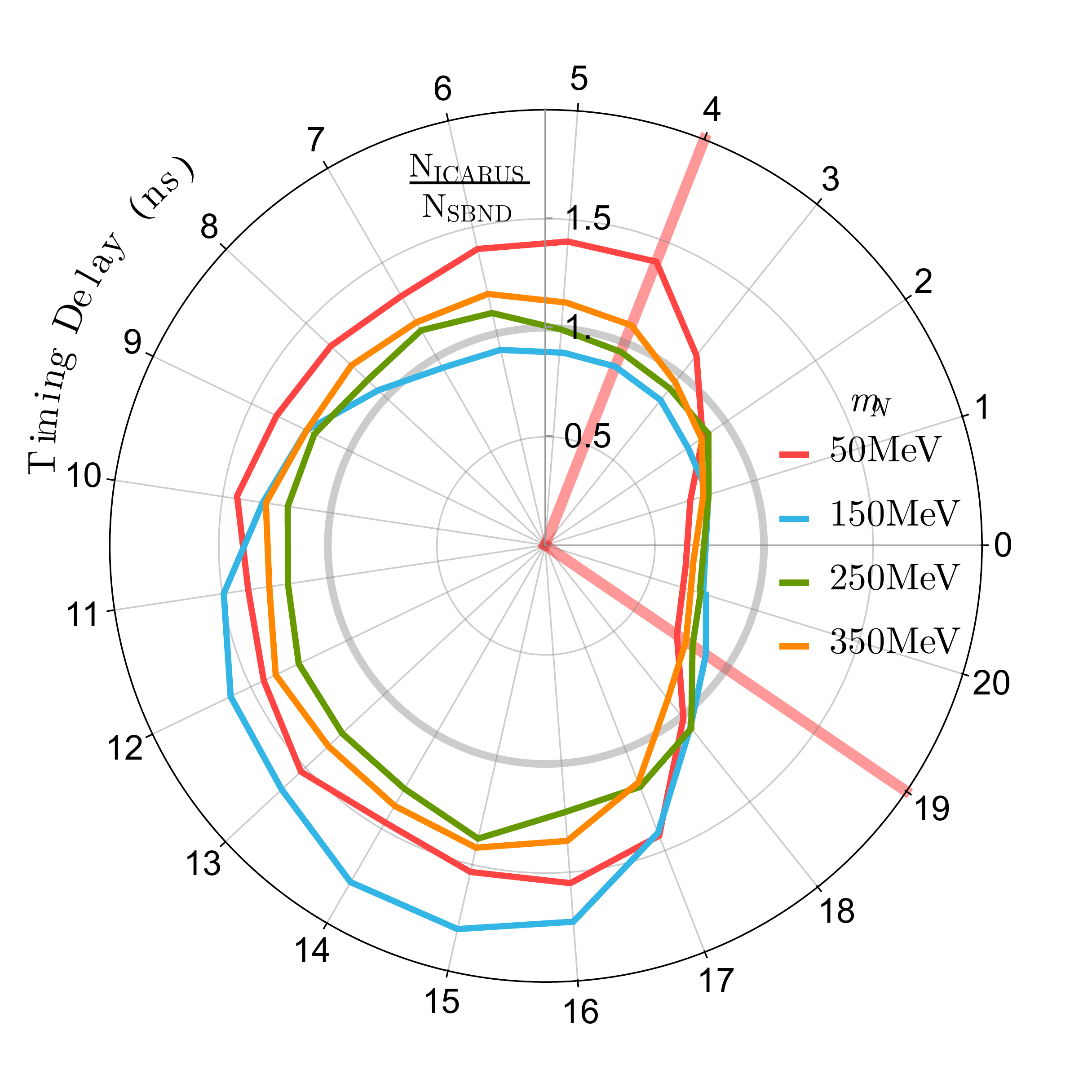}
\end{subfigure}%
~
\begin{subfigure}[t]{0.5\textwidth}
\includegraphics[width=\textwidth]{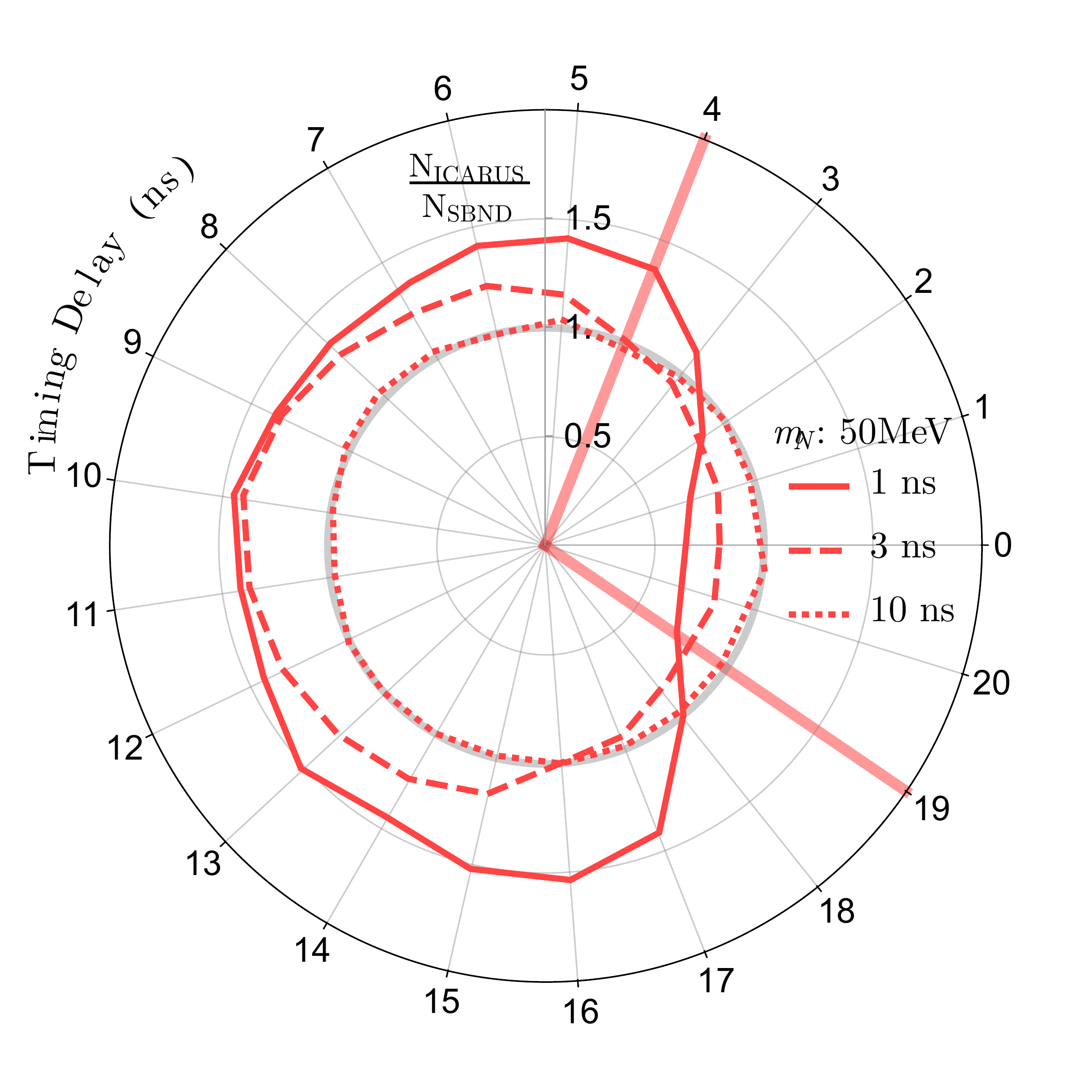}
\end{subfigure}

\caption{\label{fig:icarus_the_great} Left: Ratio of arrival time spectra of
$\ster \rightarrow \nu_\alpha e^+ e^-$ events in ICARUS to SBND after scaling
out $1/R^2$ flux dependence. If events were due to an unknown background with a
fixed time delay after the neutrino beam, one would expect the ratio to be a
constant value of 1 (shaded grey ring). As the sterile neutrinos have to travel
approximately 6 times further to ICARUS than SBND, increasingly higher energy
sterile neutrinos can leave the beam-bucket (red arc) and populate the
inter-bucket region leading to the distinct signature. This assumes a timing
resolution of $1$ ns. Right: The same as left for fixed $50$ MeV sterile
neutrino with $1$ ns (solid), $3$ ns (dashed) and $10$ ns (dotted) timing
resolution showing the decreasing effect on the ratio. }

\end{figure}

We show the consequences of this effect in \reffig{fig:icarus_the_great} where
the ratio of events at SBND and ICARUS are plotted as a function of time delay.
A constant time delay would produce a ratio of unity, and curves that lie on
the grey circle. We see a clear distortion in this ratio, with a generally low
value inside the beam-bucket window and a larger value outside. Measuring this
distortion would be definitive proof of the heavy particle having propagated
the distance from target to detector and not merely being produced \emph{in
situ}. On the right panel of  \reffig{fig:icarus_the_great}, we show how the
attainable timing resolution affects this measurement. For a resolution of $10$
ns, there is no spectral difference, but distortion starts to be apparent for
resolutions better than $1$ ns.

Assuming a positive signal is found and is identified as a heavy sterile
neutrino decay thanks to the time delay, the temporal and energy analyses could
be used to measure the heavy sterile neutrino mass.  For an arrival time delay
(behind a luminal or near luminal particle) over a distance $L$ denoted by
$\Delta T$, the mass of a sterile neutrino with an energy $E$ can be
reconstructed as 
\[ m_{\ster} = E\sqrt{1-\frac{1}{\left(1+\frac{c\Delta T}{L}\right)^2}}. \]
Exact knowledge of the deposited energy and time of flight would
be sufficient to establish the mass, but of course these data are in most cases
not available: energy and timing resolution impair the reconstruction, and many
channels have missing energy from active neutrinos in the final state.
Moreover, due to the cyclic nature of the BNB beam buckets, an observed event
could have originated from any of the previous buckets in the current spill,
and not just the one closest to the tagged event timing. As such the absolute
time of flight is not known. Only the relative timing since the last bucket,
$\Delta T$, is known and from this one can obtain up to 81 degenerate solutions
for the sterile neutrino mass. Although absolute timing information could be
found by studying the first few buckets for the onset of a signal, this would
rely on precise absolute timing measurements between source and detector, and
would also reduce the signal statistics by $\mathcal{O}$(0.01) and we do not
consider this information in the analysis.
Given these limitations, we have studied how well $m_\ster$ could be
reconstructed, using then energy and periodic time since last bucket $\Delta
T$. We have generated Monte Carlo event data tagging each event by an arrival
time, accounting for a systematic uncertainties on the time and energy
measurement. We smear the true energy to represent detector effects as
described above, and additionally smear the time of each event with a Gaussian of 
width $\sigma_T  \approx 1$ ns for SBND and ICARUS. We use the same
Monte Carlo analysis and test statistic as in the temporal analysis above,
expanded to include a binned energy spectra. The reconstructed mass is defined
as the mass which minimises the test statistic $t_m$. 

\begin{figure}[t]
\center
\begin{subfigure}[t]{0.5\textwidth}
\includegraphics[width=\textwidth]{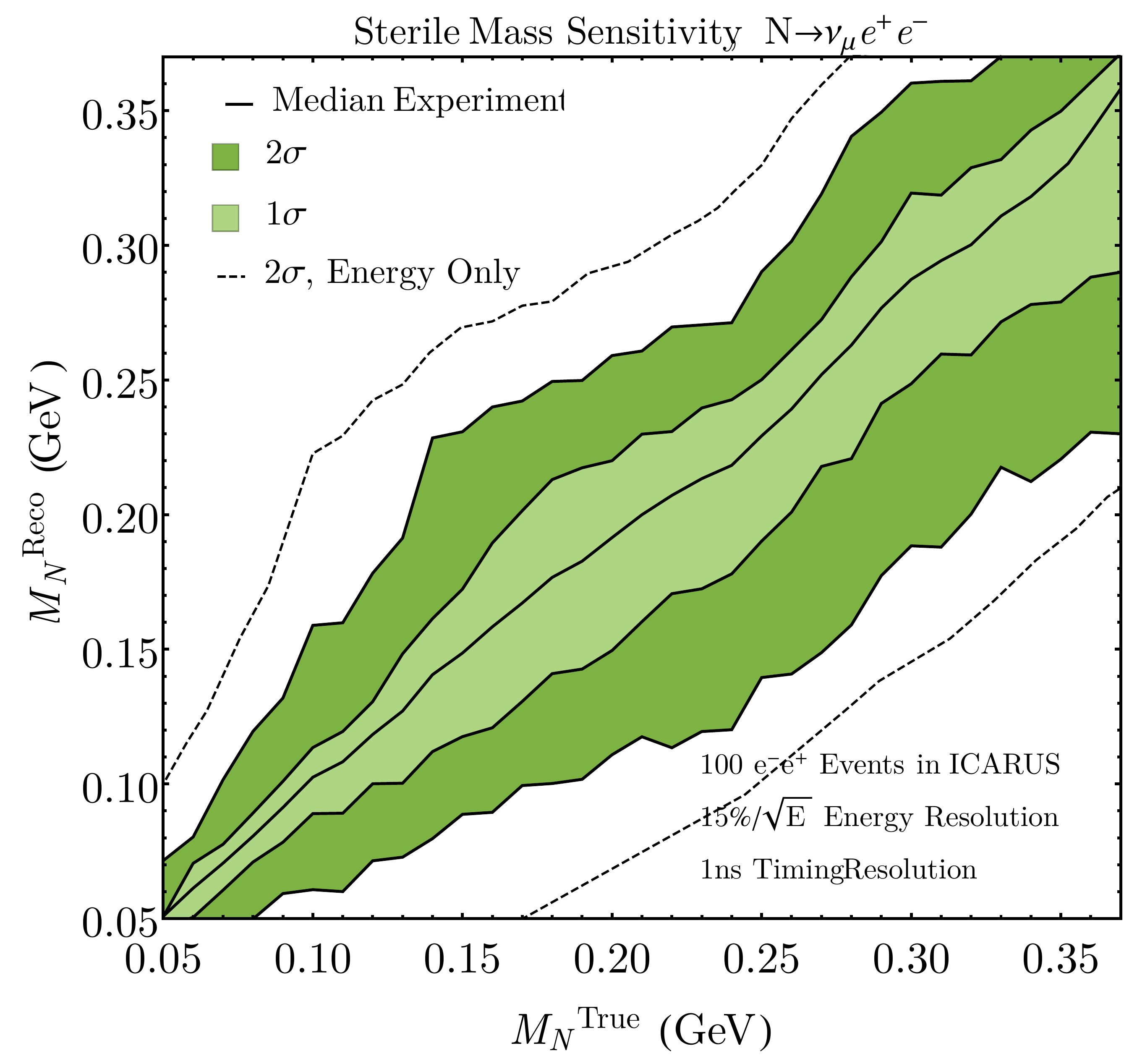}
\end{subfigure}%
~
\begin{subfigure}[t]{0.5\textwidth}
\includegraphics[width=\textwidth]{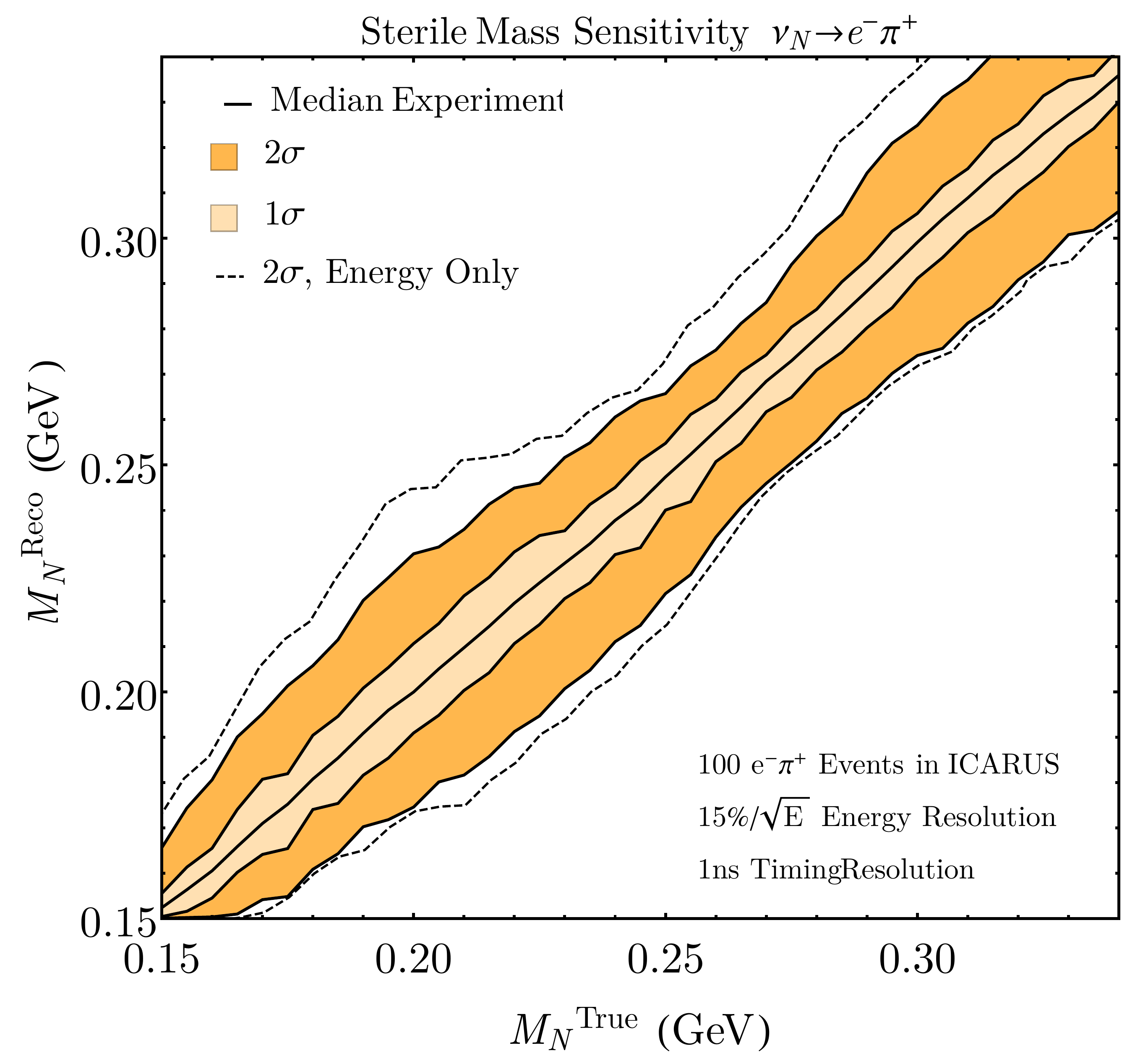}
\end{subfigure}

\caption{\label{fig:tof_scatter}The reconstructed sterile neutrino mass as a
function of true mass for energy only (dashed) and energy with timing
information (solid). The left (right) panel assumes the decay $\ster\to\nu
e^+e^-$ ($\ster \to e^\pm\pi^\mp$). Both plots assume that 100 events are seen
at ICARUS and that a 1 ns timing resolution is achieved.}

\end{figure}

The results of our analysis are shown in
\reffig{fig:tof_scatter}. In both panels, we show the allowed
region in reconstructed mass as a function of true sterile neutrino mass for an
energy only analysis (dashed black lines), as well as for an energy and
time-of-flight analysis (coloured bands). The left panel shows the results for
the fully leptonic decay $\ster \rightarrow \nu e^+ e^-$ while the right panel
shows our results for the semi-leptonic $\ster \rightarrow e^\pm \pi^\mp$
channel. In the case of the 2-body $\ster \rightarrow e^\pm \pi^\pm$ channel,
resolution of approx 45 MeV at 2$\sigma$ level is achievable for the entire
range of sterile neutrino mass allowed. We estimate the $\ster\rightarrow \mu^-
\pi^+$ channel would be approximately 10\% better due to the improved energy
resolution possible when reconstructing muons in LAr. For these
semi-leptonic decays the energy spectrum is very informative, as the parent
particle's energy can be reconstructed from the invariant mass of the decay
products' four-momenta. As such, we see temporal information only trivially
improves the reconstruction of parent mass.  In contrast, for the 3-body $\ster
\rightarrow \nu e^+ e^-$ channel, there is significant missing energy taken
away by the active neutrino. In this case, timing information is much more
valuable, almost halving the 2$\sigma$ mass range from around $300$ MeV to
$150$ MeV for widest region of parameter space.

\section{\label{sec:conclusions}Conclusions}

In this paper, we have studied the prospects for the measurement of MeV-scale
sterile neutrinos at the Fermilab Short-Baseline Neutrino program. MeV-scale
neutral states would naturally be produced in the Booster Neutrino Beam from
mixing-suppressed meson decays. To assess SBN's potential to constrain these
models, we have estimated the dominant backgrounds and signals. Thanks to
excellent particle identification and the distinctive kinematic properties of
our signal, high levels of background suppression can be expected, allowing SBN
to improve on the current bounds on decaying sterile neutrinos over most of the
parameter space. 
We have shown that, in the absence of signal, SBN can place bounds on the
active-sterile mixing-matrix elements of $|U_{e4}|\leq 10^{-6}$ for $m_\ster
\leq 33$ MeV and $|U_{\mu 4}|^2 \leq 2 \times 10^{-8}$ for $m_\ster \leq 138$
MeV in the $\ster \rightarrow \nu e^+ e^-$ channels. For semi-leptonic decays,
these bounds increase  up to $|U_{e4}|\leq 8 \times 10^{-10}$ for $m_\ster \leq
388$ MeV and up to $|U_{\mu 4}|^2 \leq 7 \times 10^{-10}$ for $m_\ster \leq 493
$ MeV. The neutral pion decay channel, $\ster \rightarrow \nu \pi^0$, which may
be the dominant decay mode for masses in the range $m_\pi^0 \leq m_\ster \leq
m_{\pi^\pm} +m_\mu$, can be used to place bounds of around $|U_{\alpha
4}|^2\leq 3 \times 10^{-9}$.
We have also discussed searches for non-minimal models of heavy sterile
neutrino decay, which could lead to observable decays over a wide range of
parameter space which is conventionally excluded if the branching ratios are
assumed to arise from the minimal model. We have shown how to map existing
minimal-model bounds onto non-minimal models and how bounds could be weakened
in the case of specific enhancements to a decay channel. This motivates the
search for particle decays in all channels over the full parameter space. We
argue that some of these decay channels considered in this work are in fact
poorly constrained by similar experiments, and show that SBN could place the
first direct bounds on these processes.

We have commented in detail on the phenomenological role of timing
information in this analysis. As well as providing a means of background
suppression, nanosecond scale timing resolution could allow SBN to make direct
measurements of the kinematic properties of heavy particle propagation. We have
shown that if 100 events are seen at ICARUS, a $3.5$ ns timing resolution would
allow an observable timing delay to be established at $3\sigma$ in $95\%$ of
experiments. We have seen that the unique design of SBN would allow for the
distribution of event times to be compared between the nearest and farthest
detectors, allowing for any model with a finite time delay to be excluded when
compared to a propagating sterile neutrino model. We have also shown how timing
information can be used in sterile neutrino mass reconstruction. For the decay
$N\to\nu e^+e^-$, the inclusion of event timing information (with an assumed 1
ns resolution) can lead to the $2\sigma$ allowed region being reduced from
around $\pm300$ MeV to $\pm150$ MeV.

We point out that this analysis is complementary to the central
physics programme of SBN --- studying eV-scale oscillating sterile neutrinos
--- and requires no additional detector or beam modifications.
We have shown that SBN could contribute valuably to the search for
sterile neutrino decays-in-flight, and moreover, if an anomalous signal is
discovered, would play a central role in determining its origin,
and the necessary extension of the SM.

\acknowledgments

We would like to thank Andrezj Szelc for his input at various stages of this
work, and also to Jonathan Asaadi for helpful discussions and encouragement at
the start of this project.

This work has been supported by the European Research Council under ERC Grant
``NuMass'' (FP7-IDEAS-ERC ENC-CG 617143), by the European Union FP7
ITN-INVISIBLES (Marie Curie Actions, PITN-GA-2011-289442), as well as from the
H2020 funded ELUSIVES ITN (H2020-MSCA-ITN-2015, GA- 2015-674896-ELUSIVES) and 
InvisiblePlus (H2020-MSCA-RISE-2015, GA-2015-690575-InvisiblesPlus).

\appendix

\section{\label{app:decayrates}Appendix: Decay rates in the minimal model}

The decay rates in our study follow the notation of \refref{Atre:2009rg}. We
repeat them here in the interests of clarity.
The dominant visible decay for sterile neutrinos with masses below the pion
mass is into an electron positron pair. The total rate can be express as
\begin{align*} \Gamma\left(\ster\to \nu_\alpha e^+e^-\right) =
\frac{G_\text{F}^2m_\ster^5}{96\pi^3}\left|U_{\alpha 4}\right|^2&\left[\left(
g_Lg_R + \delta_{\alpha e}g_R\right)I_1\left(0,\frac{m_e}{m_\ster}, \frac{
m_e}{m_\ster}\right)\right.\\ &\left.\qquad + \left(g_L^2 + g_R^2 +
\delta_{\alpha
e}(1+2g_L)\right)I_2\left(0,\frac{m_e}{m_\ster},\frac{m_e}{m_\ster}\right)\right],
\end{align*}
where $g_L = -1/2 + \sin^2\theta_\text{W}$, $g_R = \sin^2\theta_\text{W}$ and
\begin{align*} I_1(x,y,z) & =12 \int_{(x+y)^2}^{(1-z)^2}
\frac{ds}{s}(s-x^2-y^2)(1+z^2-s)\sqrt{\lambda(s,x^2,y^2)}\sqrt{\lambda(1,s,z^2)},\\
I_2(x,y,z)&
=24yz\int_{(y+z)^2}^{(1-x)^2}\frac{ds}{s}\left(1+x^2-s\right)\sqrt{\lambda\left(s,y^2,z^2\right)}\sqrt{\lambda\left(s,y^2,z^2\right)},\\
\lambda(a,b,c) &= a^2+b^2+c^2 - 2ab-2bc-2ca.  \end{align*}
The decays into a charged lepton and a pion are given by 
\[ \Gamma\left(\ster\to l^\pm\pi^\mp\right) =
\left|U_{l4}\right|^2\frac{G_\text{F}^2f_\pi^2 |V_{ud}|^2
m_\ster^3}{16\pi}I\left(\frac{m_l^2}{m_\ster^2} ,
\frac{m_\pi^2}{m_\ster^2}\right) , \] 
with \[ I(x,y) = \left[ \left( 1+x+y\right) \left(1+x\right) -4 x\right]
\lambda^\frac{1}{2}\left(1,x,y\right).  \]
For $N\to e^\pm\pi^\mp$ the kinematic function $I(x,y)$ produces only weak
suppression ($I(x,y)\geq 0.5$) for sterile  neutrino masses above
$m_\ster\gtrsim 150$ MeV, whilst for $N\to \mu^\pm\pi^\mp$ the equivalent
threshold is $m_\ster\gtrsim 270$ MeV.

\section{Appendix: Potential backgrounds \label{app:bg}} 

In all channels a cut on vertex activity is applied as described in
\refsec{sec:backgroundestimate} above. In this section, we provide a brief
description of the backgrounds and additional cuts considered for each channel.

\subsection{$\ster \rightarrow e^\pm \pi^\mp$}

The expected numbers of $e \pi$ events in the SBN detectors is significantly
smaller than that of the $\mu \pi$ channel, as the fraction of intrinsic
$\nu_e$ in the BNB beam is of $\mathcal{O}(1\%)$ level in comparison to
$\nu_\mu$. However, additional backgrounds to the $e \pi$ channel originate
from the dominant $\nu_\mu$ component of the beam. CC $\nu_\mu$ events which
contain an additional photon $(\mu+\gamma)$ have the potential to be be
mis-identified as an $(\pi e)$ event, provided the muon has a sufficiently
short track length, $<$ 0.5 m, in order to mimic a $\pi^-$. Additionally the
photon must be mis-identified as an electron, with an efficiency of 94\% using
$dE/dx$ measurement, and must convert to an $e^+e^-$ pair close enough to the
interaction vertex as so there is no visible gap, $\leq 3$ cm. As energy
resolution for EM showers is lower than muons, the invariant mass cut is less
powerful requiring all events have an invariant mass below 500 MeV.  A cut on
the opening angle between lepton on meson, $\theta_{l \pi} < 40^\circ$ as well
as individual emission angles, $\theta_{l,\pi} < 80^\circ$ further reduces the
potential background.  The $e^- \pi^+$ channel is one of the cleanest channels
under consideration in this paper, with 9,223 events in SBND reducing to 22
expected events post cuts, and with \muboone\ and ICARUS expecting a reduction
of 784 (1,317) events to 2 and 3 respectively, with a signal efficiency of
71\%.

\subsection{$\ster \rightarrow \nu_\alpha e^+ e^-$ and $\ster \rightarrow \gamma \nu_\alpha$ }

A sufficiently boosted, and thus overlapping, $e^+e^-$ pair is topologically
indistinguishable from a converted photon in a LAr detector. Additional,
non-topological measures such as the rate of energy loss, $dE/dx$, is also
identical to a pair-converted photon. Thus we split this channel into two sub
categories, when the $e^+e^-$ is overlapping and photon-like, defined to be all
events whose angular separation is $\leq 3^\circ$\cite{Spitz:2011wba} and all
remaining separable two track events. The opening angle between the $e^+e^-$ in
a photon pair production scales roughly as $\approx m_e/E_\gamma$, with
$3^\circ$ corresponding to 100 MeV and used as a lower bound on energy. These
backgrounds are also applicable to the $\ster \rightarrow \gamma
\nu_\alpha$ channel.

The predominant source of backgrounds is the decay of a neutral pion in which a
single photon is not resolved or escapes the fiducial volume. This background,
however, is relatively isotropic in distribution in stark contrast to the very
forward signal arising from the decay in flight of $N$. We place cuts on
visible photon energy, $E_\gamma \geq 300 $ MeV, and angle of the observed
photon to the beamline, $\theta_\gamma \leq 5^\circ$. This reduces the number
of expected events from 42,580 (3,620 and 6,082) to 176 (46, and 110) events in
SBND (\muboone\ and ICARUS), while retaining a signal efficiency of 93\%.

For the opposite scenario both daughter electrons have a well defined and large
separation and thus can cleanly be identified as two distinct single electron
showers. There are few significant processes that produce high energy,
distinguishable $e^+e^-$ pairs. Instead the majority of the backgrounds are due
to misidentifying two photons. We apply the same photon cuts as defined above.
To further reject backgrounds in this channel, we apply a cut on the angle of
separation between the distinct $e^+e^-$ tracks of $\theta_\text{sep}\leq 40
^\circ$ and total energy, $E_{e^+}+E_{e^-} \geq 100$ MeV. This reduces the
number of expected background events from 173 to 5 for SBND.

\subsection{$\ster\rightarrow \pi^0 \nu_\alpha$} 

Single neutral pions are
produced in great numbers at the three SBN detectors, so the lack of any
nuclear recoil is crucial in eliminating the incoherent neutral pion production
background. Only events in which two photons convert inside the fiducial volume
and reconstruct the pion invariant mass are accepted. We further require the
reconstructed pion is within $10^\circ$ of the beamline and has an energy above
500 MeV. SBND expects 127,211 $\pi^0$ events, of which $\approx 602$ survive
all cuts with a signal efficiency of 32\% for a sample 350 MeV sterile
neutrino. \muboone\ (ICARUS) sees a similar reduction, from 10,813 (18,172)
events to 51 (86) post cuts.
 
\subsection{$\ster \rightarrow \nu \mu^\pm \mu^\mp$ }

The primary background for this channel is genuine $\nu_\mu$ CC events in which
a $\pi^\pm$ is also produced and is misidentified as a secondary muon. All
pions with reconstructed tracks longer than 50 cm are considered a potential
muon. After this track length cut, there is 2,044,380, 177,972 and 292,034
events in SBND, \muboone~ and ICARUS respectively. As we cannot directly
reconstruct the parent sterile neutrino mass or angle, we again rely on the
kinematical difference between scattering events and decays. After these cuts,
significant backgrounds remain, and we use a multivariate analysis, an adaptive
boosted decision tree (BDT), in order to further reduce them. We use five
parameters in this analysis, the energy and angle with respect to the beamline
of each muon, as well as the angular separation between both muons. We take a
minimum muon energy of 200 MeV. Cutting on the BDT response variable allows
for background efficiency of 0.13\%, with a corresponding signal efficiency of
44\%. This allows for a $S/\sqrt{S+B} \approx 8$ with approximately 1000
sterile neutrino events.  Similar performance is achievable at \muboone\ and
ICARUS, with 117,972 and 292,034 events being reduced to 534 and 876 events
respectively.

\subsection{$\ster \rightarrow \nu e^\pm \mu^\mp$ }

We consider here two potential sources of backgrounds: the first derives from
true $\nu_\mu$ CC events in which a single photon, either from nuclear
processes or from the decay of a $\pi^0$ in which only photon converts inside
the fiducial volume, subsequently mimics the electron. We apply the same cuts
on the photon as in previous channels. Secondly we consider intrinsic $\nu_e$
CC events in which a final state $\pi^\pm$ is misidentified as a muon due to a
long ($\geq 50$ cm) track in the TPC. In conjunction with the requirement of no
visible scattering vertex we expect 7,103, 618 and 1,014 events in SBND,
\muboone\ and ICARUS, respectively. To reduce this further we employ the same
multivariate analysis as described for the $\ster \rightarrow \nu \mu \mu$
channel above, assuming a representative 250 MeV sterile neutrino decaying. A
cut on the BDT allows for a background efficiency of 0.5\% , signal efficiency
of 36\% with a resultant $S/\sqrt{S+B}$ of 7.9. For \muboone\ and ICARUS the
backgrounds, 618 and 1,014 respectively, can be brought down to sub 10 events.

\subsection{Non-Beam related backgrounds}

Cosmogenic events are a potential source of background for any analysis at SBN.
In the case of cosmic muons, \icarus\ expects to see approximately $2.5 \times
10^{6}$ cosmic events in the 211 second beam spill, which are reduced to
approximately 5 events expected after utilising the spill structure,
scintillation light patterns and cuts on $\frac{d E}{d x}$
\cite{Antonello:2015lea}.  Alongside this impressive cosmic rejection, our
signal events are focused heavily along the beamline, hence we do not expect
cosmics to be a major source of background to any channel. In situ beam-off
cosmic studies will also allow potential backgrounds to be extremely well
understood by the time of an analysis such as this, and for these reasons, we
do not include cosmogenic backgrounds in our analysis. 

\section{PS-191 Bound Reproduction\label{sec:ps191}}

As a consistency check of our methodology we reproduce here the bounds on
$|U_{e4}|$ and $|U_{\mu 4}|$ for sterile masses below $m_\pi$ as published by
PS-191. The detector geometry is assumed to be $6\text{m} \times 3\text{m}
\times 12 \text{m}$ and was located 128m downstream of the Beryllium target
using 19.2 GeV protons from the PS proton beam.  Fluxes of all neutrinos
produced from pion sources at PS-191 were obtained from \cite{ps191THesis}. No
accurate kaon sources could be obtained and as such only low mass bounds are
reproduced here. It must be noted that PS-191 ignored all neutral current
contributions to $\ster \rightarrow \nu_\alpha e^+ e^-$ and assumed the sterile
neutrinos were Dirac particles; the effect of this is that the bounds published
are not directly comparable to the minimal model discussed above, and must be
scaled appropriately. The bounds reproduced are in good agreement with published
data.

\begin{figure}
\centering
\includegraphics[width=0.5\textwidth]{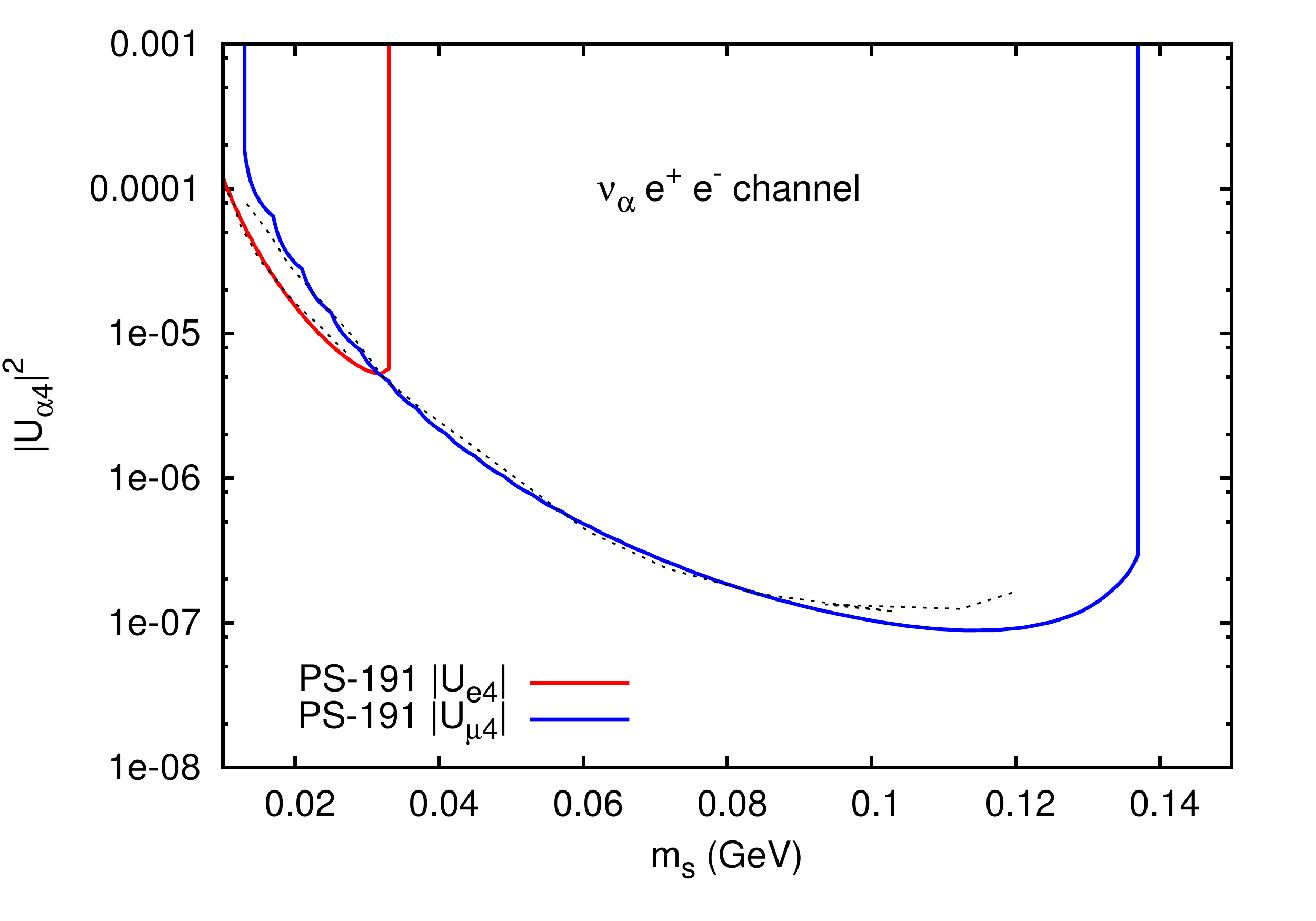}
\caption{\label{fig:ps191test} Estimated bounds on $|U_{e4}|^2$ and $|U_{\mu
4}|^2$ for a Dirac heavy sterile neutrino decaying to $\nu_\alpha e^+ e^-$ at
PS-191. The dotted black lines are the 90\% CL results as published by PS-191,
and the blue and red curves are the results of our simulation for $0.86 \times
10^{19}$ POT.}
\end{figure}


\bibliographystyle{apsrev4-1}
\bibliography{lib}{}

\end{document}